\documentclass[11pt]{article}
\usepackage{graphicx}
\usepackage{natbib}
\usepackage{xspace}
\bibpunct{(}{)}{;}{a}{}{,} % to follow the A&A style
\newcommand{\captionfonts}{\small}

\makeatletter  % Allow the use of @ in command names
\long\def\@makecaption#1#2{%
  \vskip\abovecaptionskip
  \sbox\@tempboxa{{\captionfonts #1: #2}}%
  \ifdim \wd\@tempboxa >\hsize
    {\captionfonts #1: #2\par}
  \else
    \hbox to\hsize{\hfil\box\@tempboxa\hfil}%
  \fi
  \vskip\belowcaptionskip}
\makeatother   % Cancel the effect of \makeatletter

\newcommand\nifsx{${}^{56}$Ni\xspace}

\def\inst#1{$^{#1}$}

\begin{document}

\title{Supernova Explosions inside Carbon-Oxygen Circumstellar Shells}
%\titlerunning{Supernovae in C--O shells}
   \author{S.I. Blinnikov \inst{1,2,3},  E.I. Sorokina  \inst{2,3} \\
\\
% }
%
%   \institute{
% \and
\inst{1}Institute for Theoretical and Experimental Physics, 117218 Moscow, Russia \\
% B.~Cheremushkinskaya St.~25, 117218 Moscow, Russia \\
% \and
\inst{2}Sternberg Astronomical Institute, 119992 Moscow, Russia \\
% \and
\inst{3}Max-Planck-Institut f\"ur Astrophysik, D-85740 Garching, Germany\\
% Karl-Schwarzschild-Strasse 1, Postfach 1523, D-85740 Garching, Germany
         }

% \authorrunning{Blinnikov and Sorokina}
% \date{\Large \it Crude draft, ver.1, \today }
% \date{\today }
\date{}
% \date{Received ...; accepted ... }

% \par\medskip\noindent
%

%%%%%%%%%%%OLD TEX
%\def\giveeqname{0}
%%%%%%%%%%%%%%%%%%%%%\input star3bpr
%
%\summary

\maketitle

%\begin{abstract}

\abstract{
Motivated by a recent discovery of Supernova~2010gx and
numerical results of Fryer et al.~(2010), we simulate
light curves for several type~I supernova models, enshrouded
by dense circumstellar shells, or ``super-wind'', rich in carbon and oxygen and
having no hydrogen.
We demonstrate that the most luminous events like SN~2010gx can be
explained by those models at moderate explosion energies $\sim (2 \div 3)$~foe if the total
mass of SN ejecta and a shell is  $\sim (3 \div 5)\, M_\odot $ and the radius
of the shell is  $\sim 10^{16} $~cm.
}
%\end{abstract}

% \keywords{ Stars: -- supernovae: general -- supernovae: light curves
%                }

% \maketitle

\section{Introduction}

Recently, \citet{Fryer2010} have presented arguments in favour of
type~Ia supernova (SN) explosions taking place within rather dense
extended C--O envelopes formed during a merging process in Double Degenerate (DD) scenario.
They obtained light curves which are very powerful in hard spectral range and
last sometimes very long in visible light.
On the other hand, there are observations of a very luminous type~Ic SN~2010gx
\citep{Pastorello2010},
and it was suggested that they may shine due to a shock wave propagating
in C--O-rich circumstellar material.

Motivated by these results we construct  light curves for
several non-evolutionary type~I supernova models which have dense
C--O shells or winds around them.

\section{Presupernova models}
% \subsection{If needed }
\label{sec:presn}

We have not tried to compute evolutionary or hydrodynamic models
leading to formation of structures discussed by  \citet{Fryer2010}.
All our presupernova models are constructed artificially  as described
elsewhere \citep{ChugaiEa04,BaklEa05}.
Their main parameters are given in Table~\ref{table:1}.
The interior part is a quasipolytropic model in mechanical equilibrium
where temperature is related to density as $T \propto \rho^{0.31}$.
This part has mass $M_{\rm ej}$ and radius $R_{\rm ej}$,
the subscript ``ej'' denotes here that this part becomes a supernova ejecta
after explosion.
In case of type Ib/c SN simulations  $M_{\rm ej}$ can be much less than
the total mass of the collapsing core and the condition of mechanical equilibrium
is not necessary, it is just a convenient form of parameterization of models.
The outer parts have power-law tails, $\rho \propto r^{-p}$, with $p$ given
in the Table~\ref{table:1}.
$M_{\rm Ni}$ denotes the mass of radioactive \nifsx in ejecta, and $M_{\rm w}$,
$R_{\rm w}$ are mass and radius of the ``wind''.
No attempt is done to keep equilibrium in the ``wind'', but the dynamical times
of those huge envelopes are so large that no appreciable motion has developed
during the time of light curve simulation.

The first set of models tries to mimic what is described (though very tersely)
in  \citet{Fryer2010}.
\citet{Fryer2010} claim that a double degenerate (DD) merger event leads to a density profile
like $\rho \propto r^{-4}$ which is shown in their Fig.~5,
as well as a $\rho \propto r^{-3}$ variant.
However, this structure extends only up to $r \sim 3 R_\odot$ there, while
they see shock interaction up to  $r \sim 10^{16}\mbox{cm} \sim 10^{5} R_\odot$
in their Fig.~10 \citep{Fryer2010}.
It is hard to imagine the formation of such an extended structure on a dynamical time-scale of
DD event.
Nevertheless, it is interesting to consider those extended structures independently
of DD scenario because they can help to explain extremely powerful supernovae like
SN~2010gx \citep{Pastorello2010}.
We say a few words on other feasible ways to formation of these dense circumstellar
shells in Sec.~\ref{sec:summary}.

\begin{figure}
\centering
\includegraphics[width=0.45\linewidth]{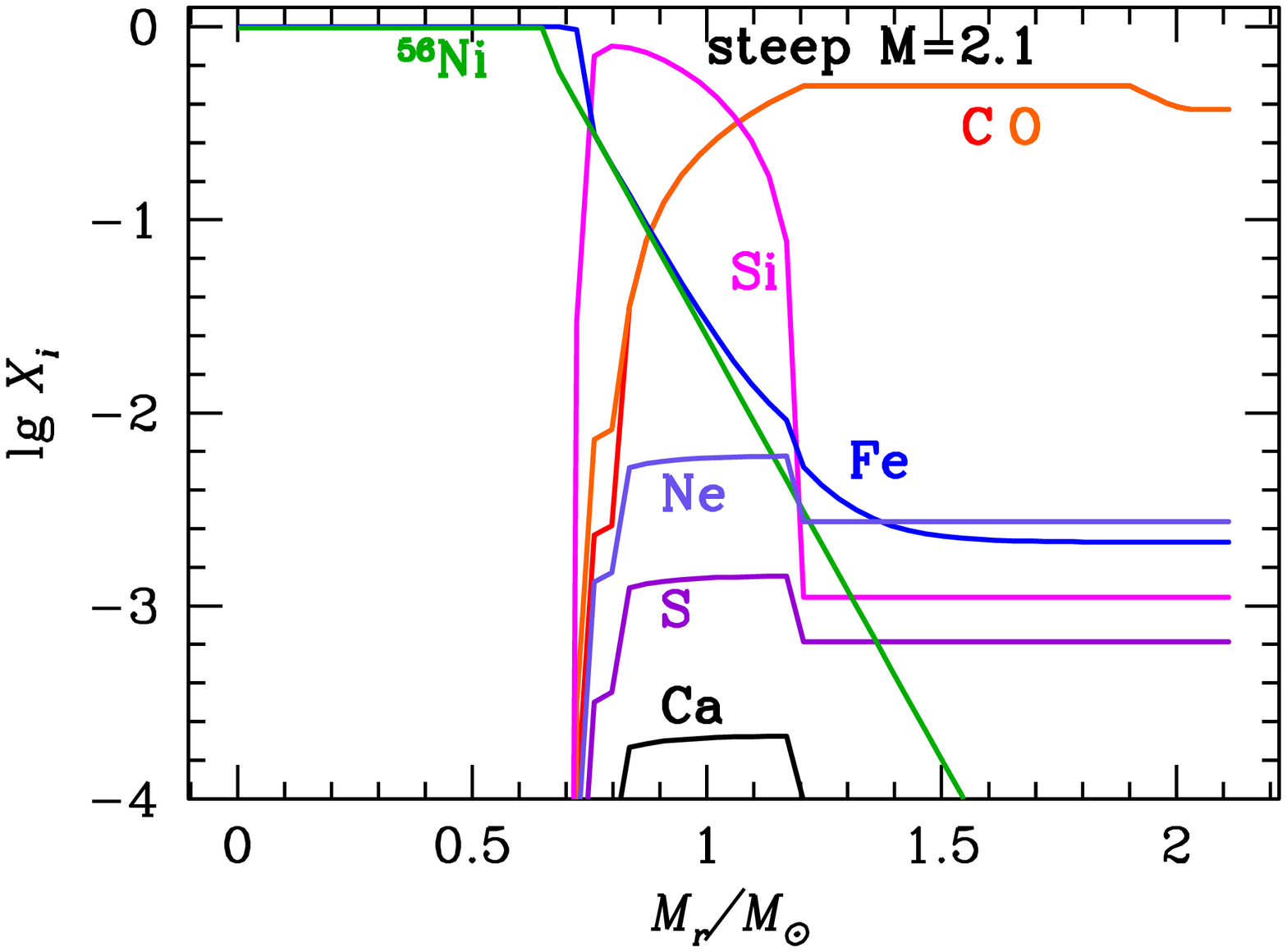}
\hskip 5mm
\includegraphics[width=0.45\linewidth]{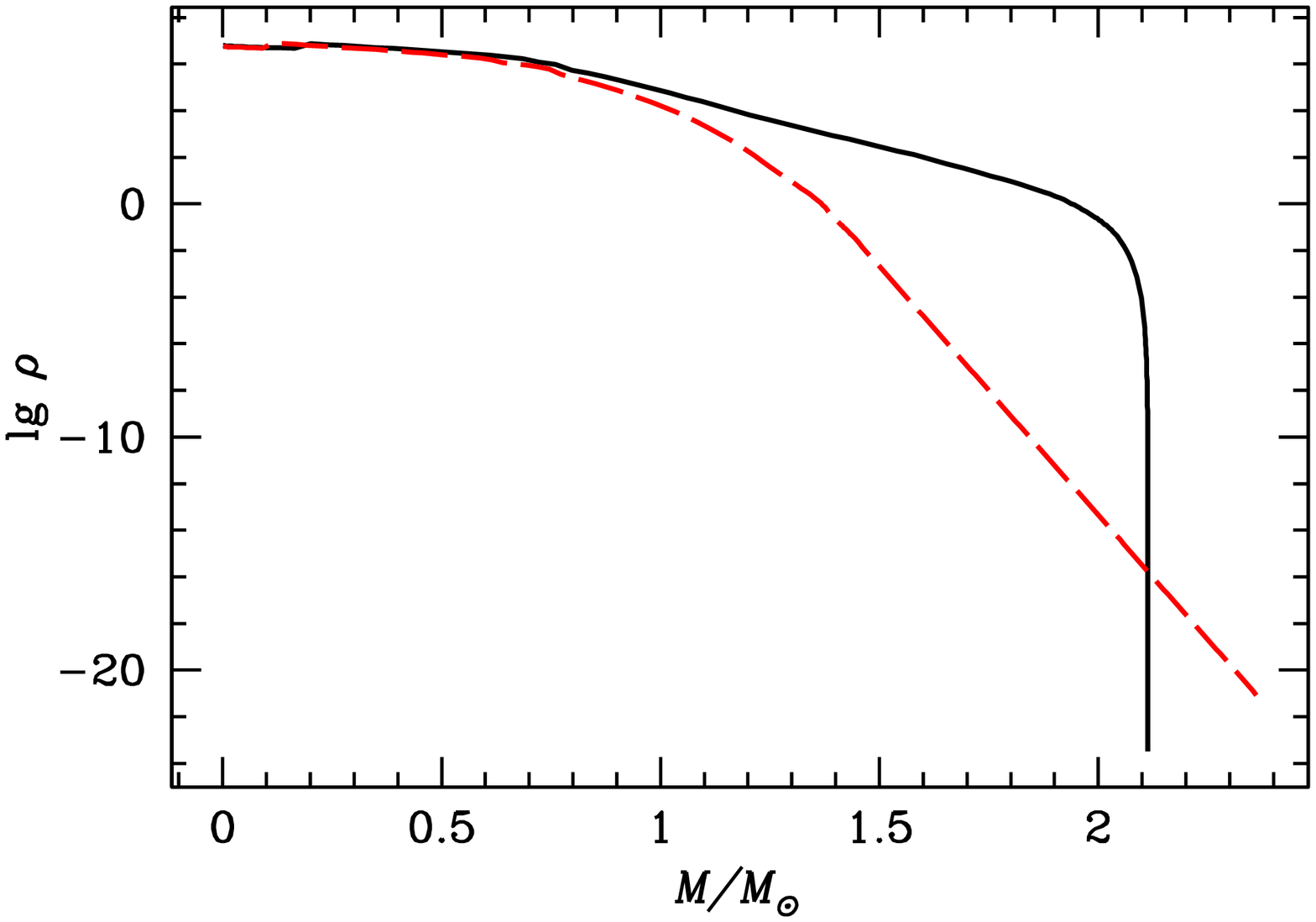}
\caption{\noindent Left: typical composition of ``Ia'' models
used in simulations. ``Fe'' includes ``Ni'' (which is not only \nifsx ) and other
iron peak elements. ``C'' and ``O'' lines almost coincide.
Right: density profiles as a function of  $M_r$ for {\tt steepIa}
 ($\rho \propto r^{-4}$; solid)
and {\tt medium bigIa} ($\rho \propto r^{-3}$; dashed) models. % $R=3\cdot 10^5 R_\odot$) models.
        }
\label{chemfry4M2}
\end{figure}

We have two types of distribution of chemical composition in our simulations.
A typical composition of the the first type is shown in Fig.~\ref{chemfry4M2}.
It tries to mimic a distribution of elements which can be produced in a thermonuclear
explosion like in SN~Ia. The models with this composition have a suffix ``Ia'' in
their names.

The second type, ``Ib'', has just a uniform distribution of elements in C--O envelope
like in outermost layers of ``Ia'' model (in left panel of Fig.~\ref{chemfry4M2}).

Density profiles for models with the ``wind'' $\rho(r)\propto r^{-4}$ and
$\rho(r)\propto r^{-3}$ are shown in  Fig.~\ref{chemfry4M2}. %  %--\ref{rhofryr3Rw5}.
All models, except for {\tt medium bigIa},
have  $T = 10^3$~K in the ``wind'', and  {\tt medium bigIa} has $T = 2.5\cdot 10^3$~K there.
Already at $T = 2.5\cdot 10^3$~K
we have got a spurious flash of light emitted by the huge envelope. % (denoted like ``wind'').
\begin{figure}
\centering
\includegraphics[width=0.45\linewidth]{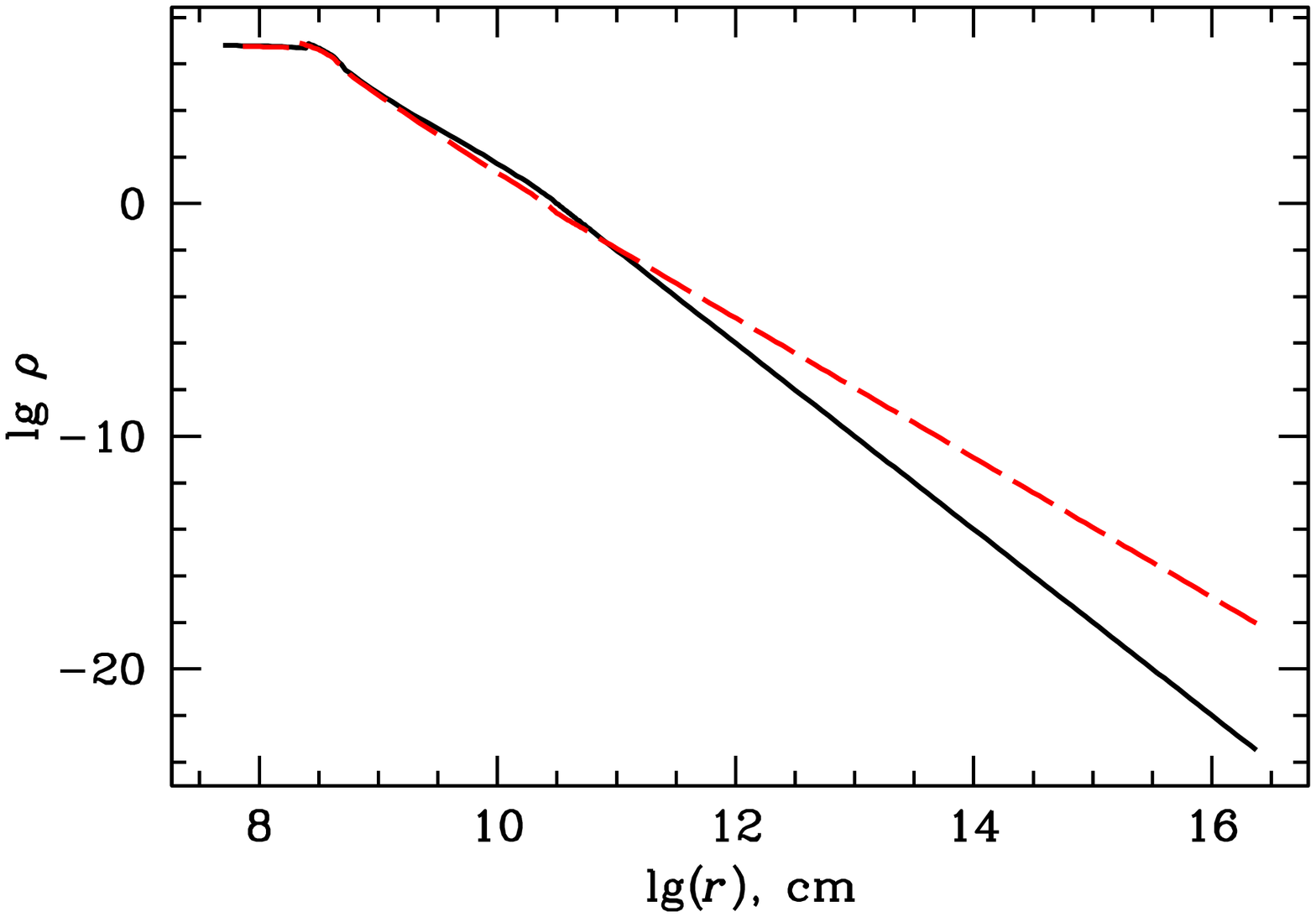}
\includegraphics[width=0.45\linewidth]{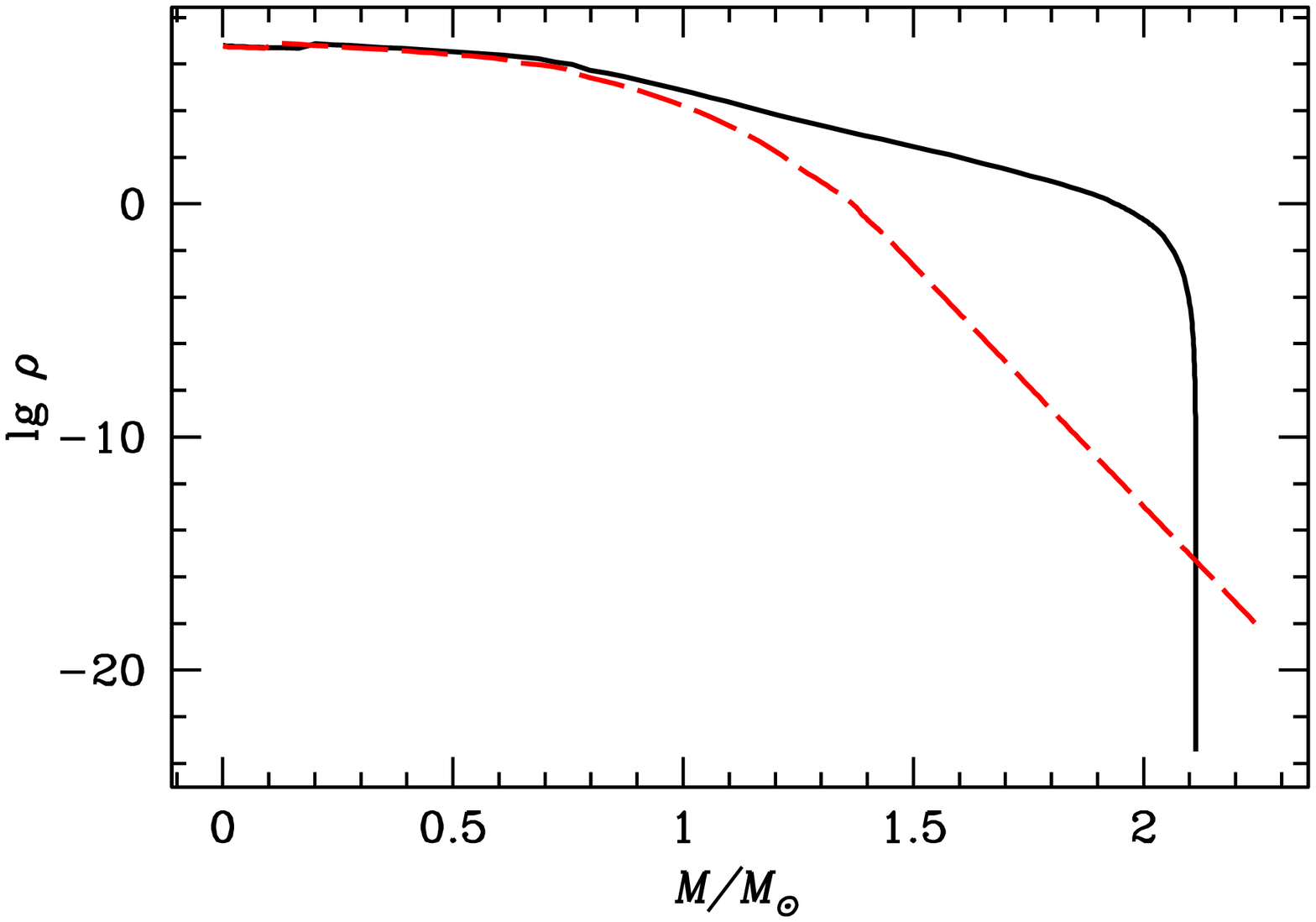}
\caption{\noindent  Left: density profiles as a function of $r$  for {\tt steepIa} (solid) and
{\tt mediumIa} (dashed) models.
(Model {\tt medium bigIa} has the same structure of the ``wind''
like the latter, $\rho \propto r^{-3}$, but its outer radius is larger,
$R_{\rm w}=3\cdot 10^6 R_\odot$).
Right: density profiles as a function of $M_r$ for the same models.
 }
\label{rhofryr4M2}
\end{figure}

% \begin{figure}
% \centering
% \includegraphics[width=0.45\linewidth]{rhorfryer3z2}
% \includegraphics[width=0.45\linewidth]{rhomfryer3z2}
% \caption{\noindent  Density profiles as a function of $r$ (left) and of $M_r$
% (right) of the model with  $\rho(r)\propto r^{-3}$ in outer layers
% with a larger outer cut of the wind ($R=3\cdot 10^6 R_\odot$).
%         }
% \label{rhofryr3z2}
% \end{figure}

% \begin{figure}
% \centering
% \includegraphics[width=0.45\linewidth]{rhorfryer3Rw5}
% \includegraphics[width=0.45\linewidth]{rhomfryer3Rw5}
% \caption{\noindent  Density profiles as a function of $r$ (left) and of $M_r$
% (right) of the model with  $\rho(r)\propto r^{-3}$ in outer layers
% with a smaller outer cut of the wind ($R=3\cdot 10^5 R_\odot$).
%         }
% \label{rhofryr3Rw5}
% \end{figure}

\begin{table}
\caption{Models (all masses $M$ and radii $R$ are in solar units)}  % title of Table
\label{table:1}      % is used to refer this table in the text
\centering                        % used for centering table
\begin{tabular}{l c c c c c c c}  % centered columns (put according each columns)
\hline %\hline                 %% inserts double horizontal lines
Model&$M_{\rm ej}$&$R_{\rm ej}$&$M_{\rm Ni}$&$p$&$M_{\rm w}$&$R_{\rm w}$&$E$, foe \\    % table heading
\hline                        % inserts single horizontal line
  {\tt steepIa}     & 2    &  1  & 0.7 & 4   & 0.1 & $3\cdot 10^5$ & 1.6 \\ % fryer4M2      % inserting body of the table
  {\tt medium bigIa}& 1.4  &  1  & 0.7 & 3   & 0.9 & $3\cdot 10^6$ & 1.6 \\ % fryer3z2
  {\tt mediumIa}    & 1.4  &  1  & 0.7 & 3   & 0.8 & $3\cdot 10^5$ & 1.6 \\ % fryer3Rw5
  {\tt shallowIb}   & 1    &  10 & 0   & 2.5 & 2.9 & $10^5$        & 3 \\ %out15p25
  {\tt standardIb}  & 0.2  &  10 & 0   & 2   & 3.5 & $8\cdot 10^4$ &  3 \\ %out24p2
  {\tt brightIb}    & 0.2  &  10 & 0   & 1.8 & 4.8 & $9\cdot 10^4$ & 1, 2, 4 \\ %out26p18z3
\hline                                   %inserts single line
\end{tabular}
\end{table}

\section{Light curve simulations}
\label{sec:simul}

We  perform calculations of the synthetic light curves
using our multi-group radiation hydrodynamic code {\sc stella} in its
standard setup \citep{BliEas1998,BliSor2004,BaklEa05,BliRop2006}.

The explosions have been simulated for all ``Ia'' variants by a ``thermal bomb'' with energy
$1.6\cdot 10^{51}$~ergs~$=1.6$~foe like in \citet{Fryer2010} .
A similar ``thermal bomb'' was used for ``Ib'' runs
(the burst duration is 10~seconds in the innermost layers of ejecta,  $\Delta M = 0.06 M_\odot$)
with variable energy $E$, see Table~\ref{table:1}.
In radiation hydrodynamics runs we used 200 (100 in the ejecta plus 100 in the
``wind'') radial mesh zones for ``Ia'' runs and
300 ones (150 plus 150) in  ``Ib'' runs.
All runs employed 100 frequency group in transport solver and relatively short
spectral line list ($\sim 1.5\cdot 10^5 $) in the opacity routine.

\section{Results}
\label{sec:result}

We report the numerical results for a set of selected models below.

\subsection{Type Ia explosions within C--O shells}
\label{ssec:Ia}

\begin{figure}
\centering
\includegraphics[width=0.45\linewidth]{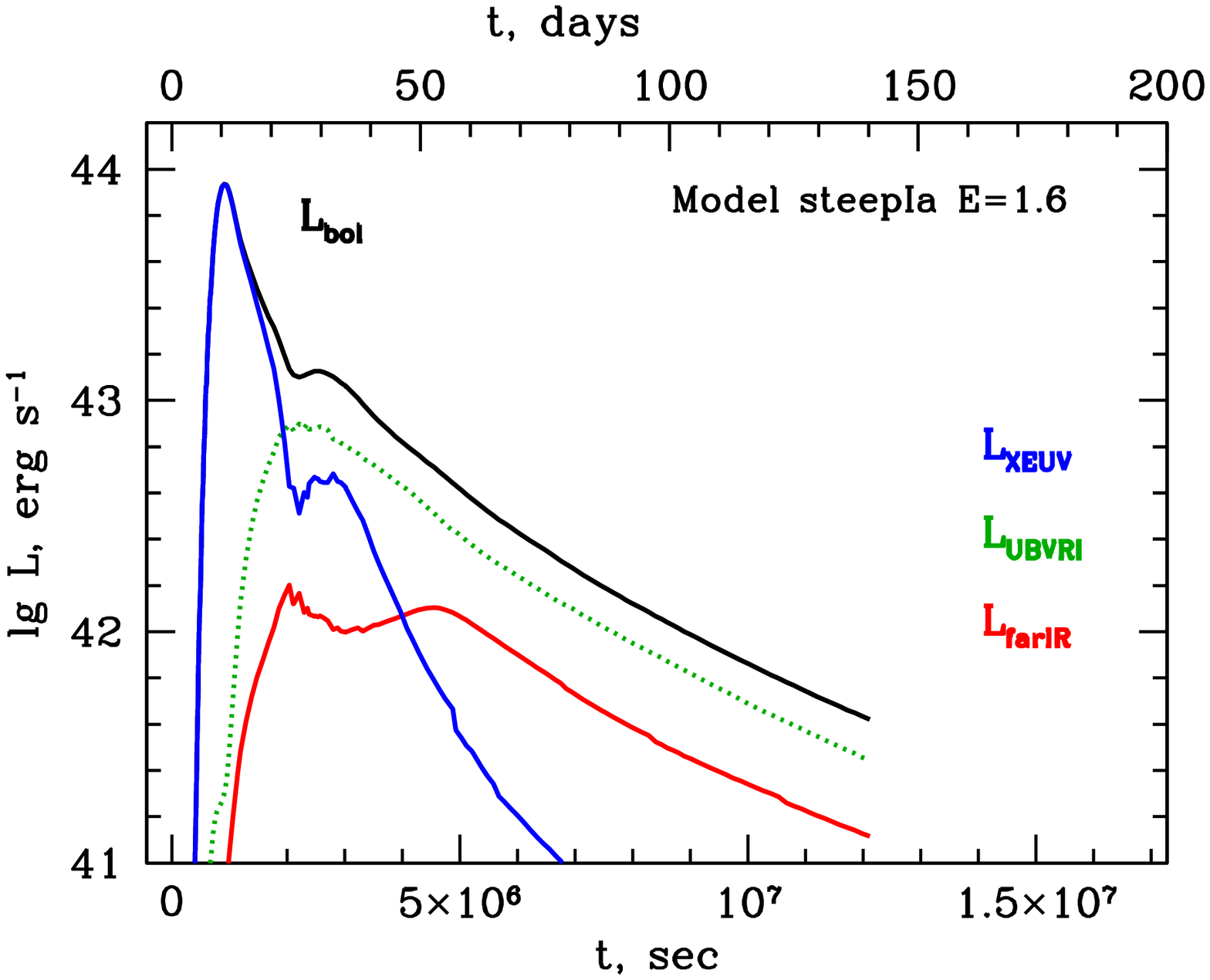}
\hskip 5mm
\includegraphics[width=0.45\linewidth]{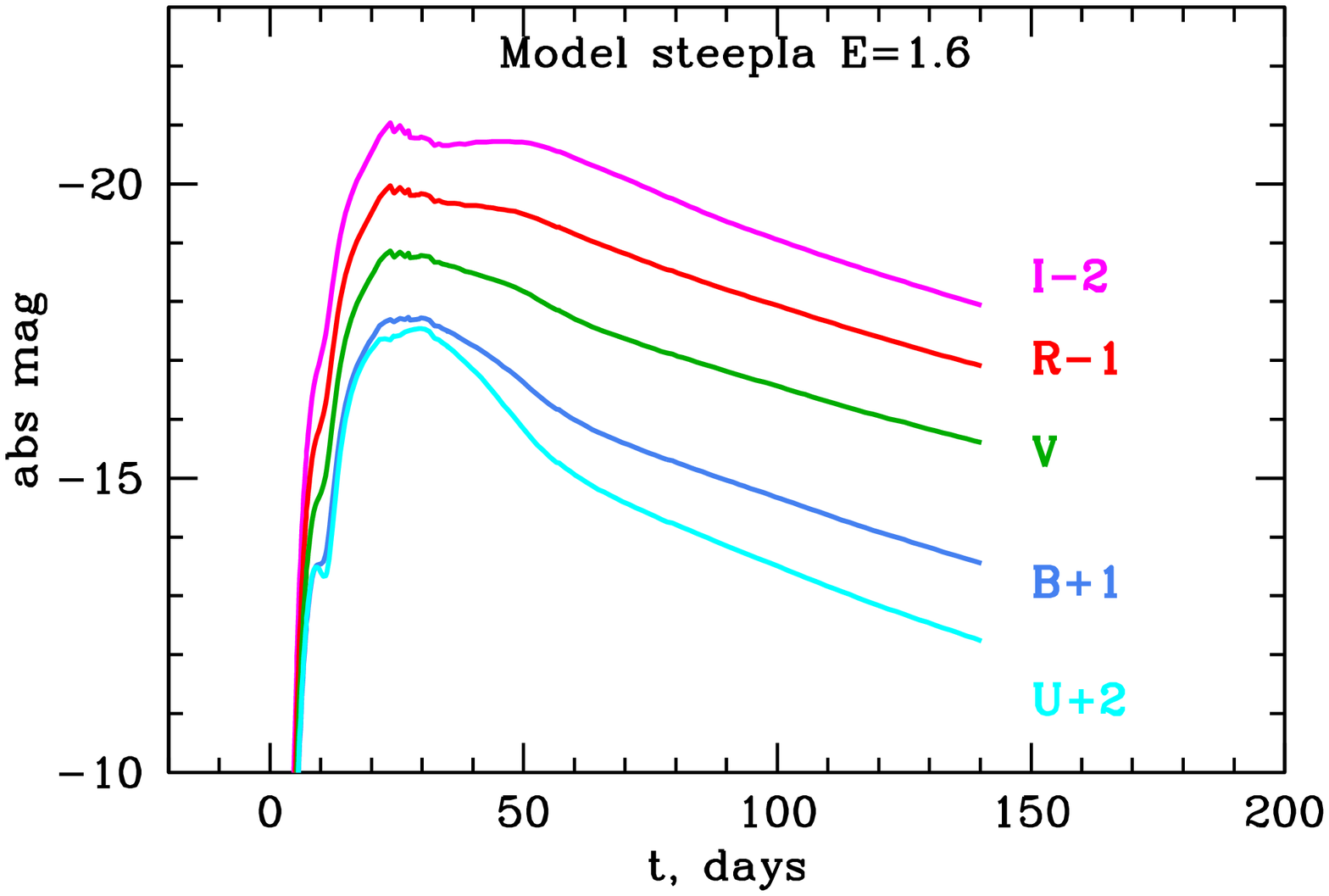}
\caption{\noindent Light curves
of the model {\tt steepIa} with  $\rho(r)\propto r^{-4}$ in outer layers.
        }
\label{LCfryr4M2}
\end{figure}

\begin{figure}
\centering
\includegraphics[width=0.45\linewidth]{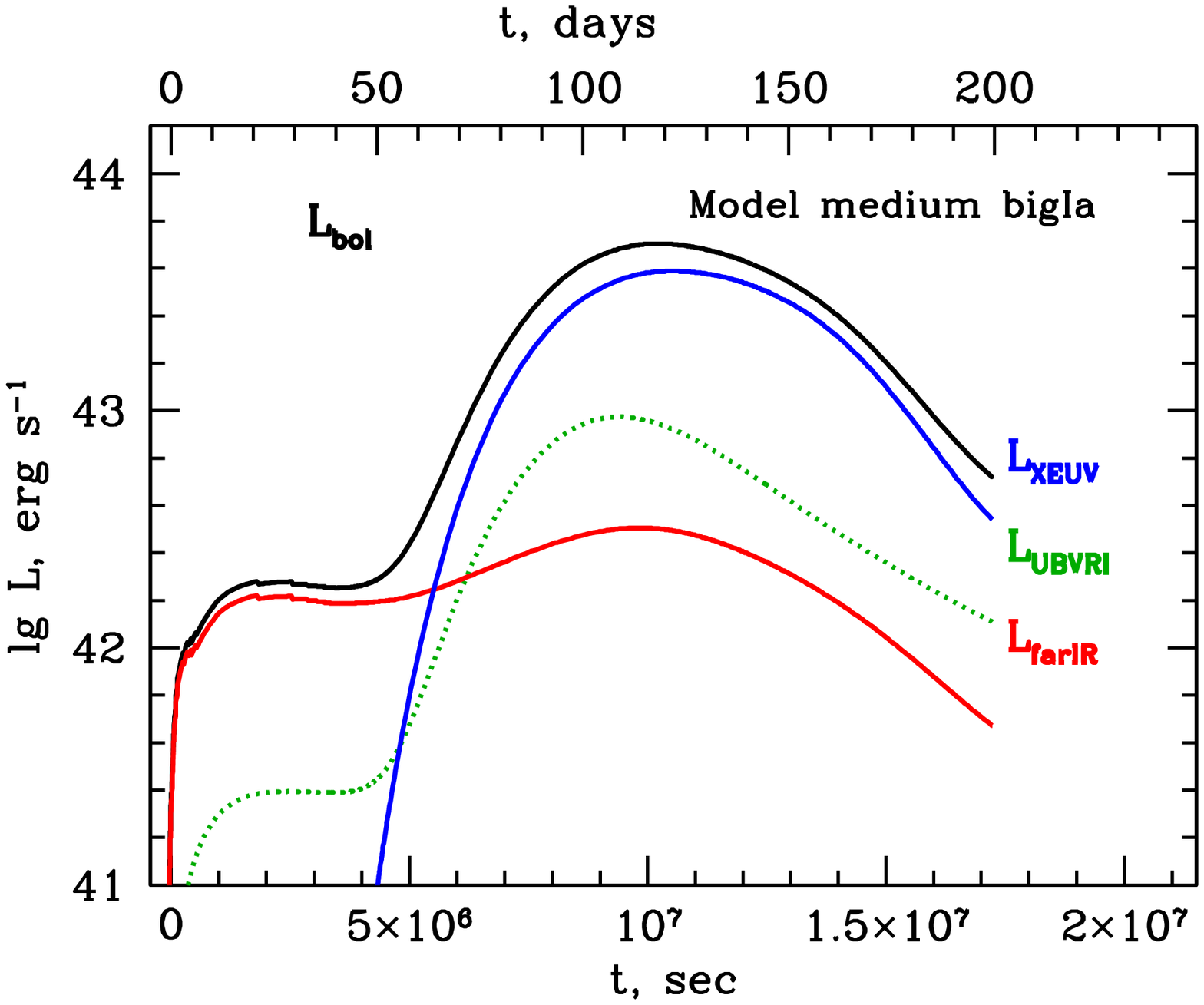}
\hskip 5mm
\includegraphics[width=0.45\linewidth]{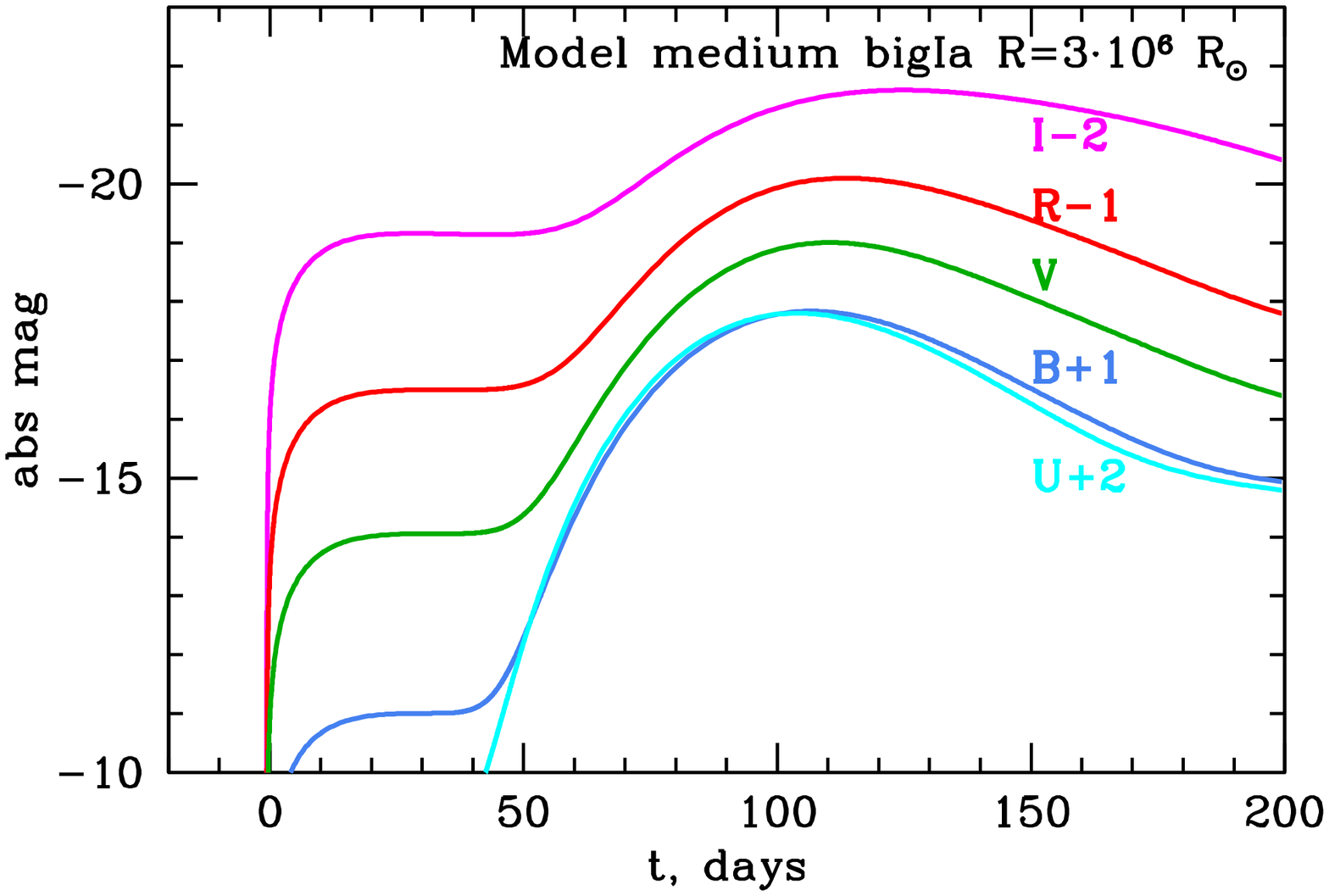}
\caption{\noindent Light curves
of the model {\tt medium bigIa} with  $\rho(r)\propto r^{-3}$ in outer layers
with a larger outer cut of the wind ($R=3\cdot 10^6 R_\odot$).
        }
\label{LCfryr3z2}
\end{figure}

\begin{figure}
\centering
\includegraphics[width=0.45\linewidth]{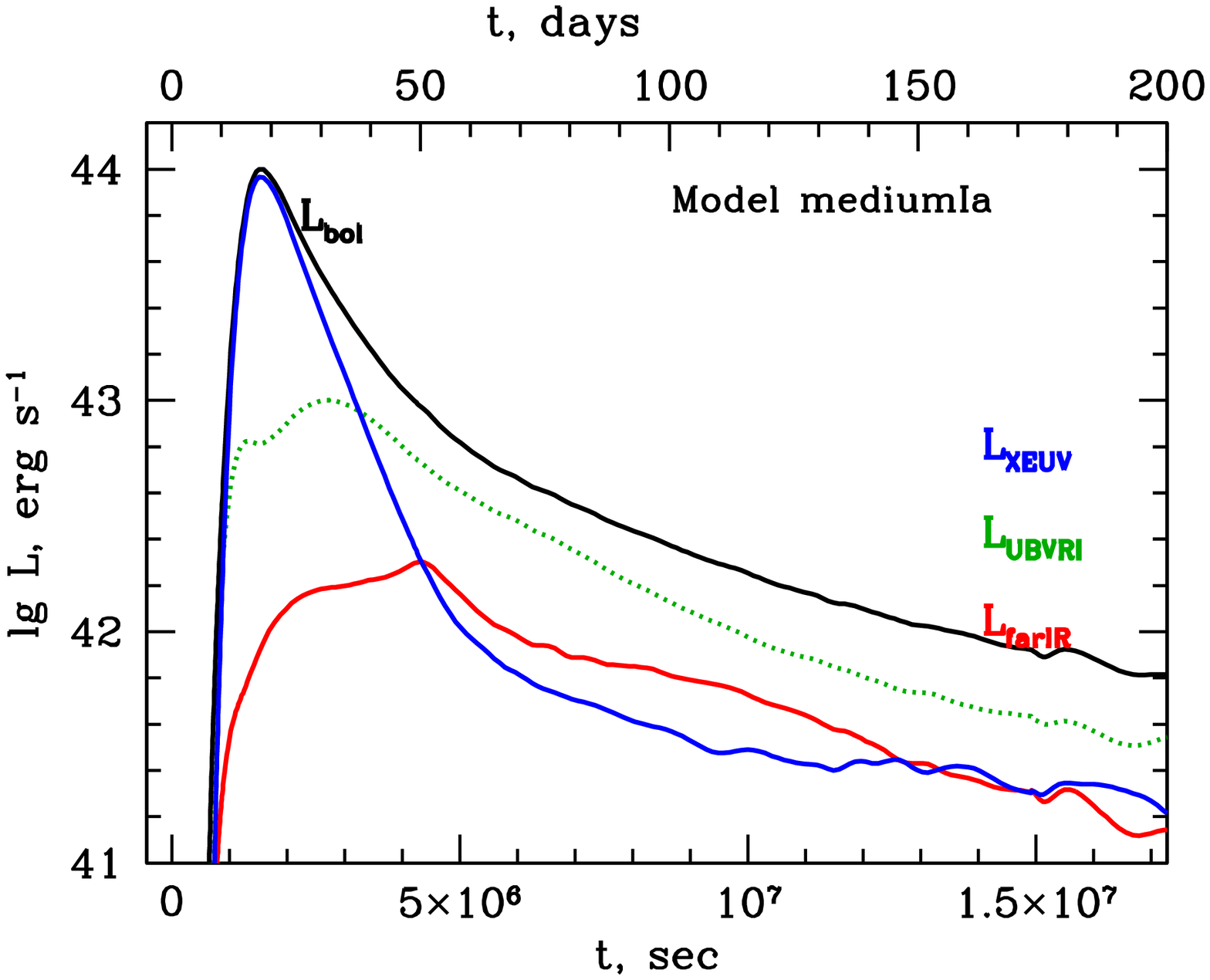}
\hskip 5mm
\includegraphics[width=0.45\linewidth]{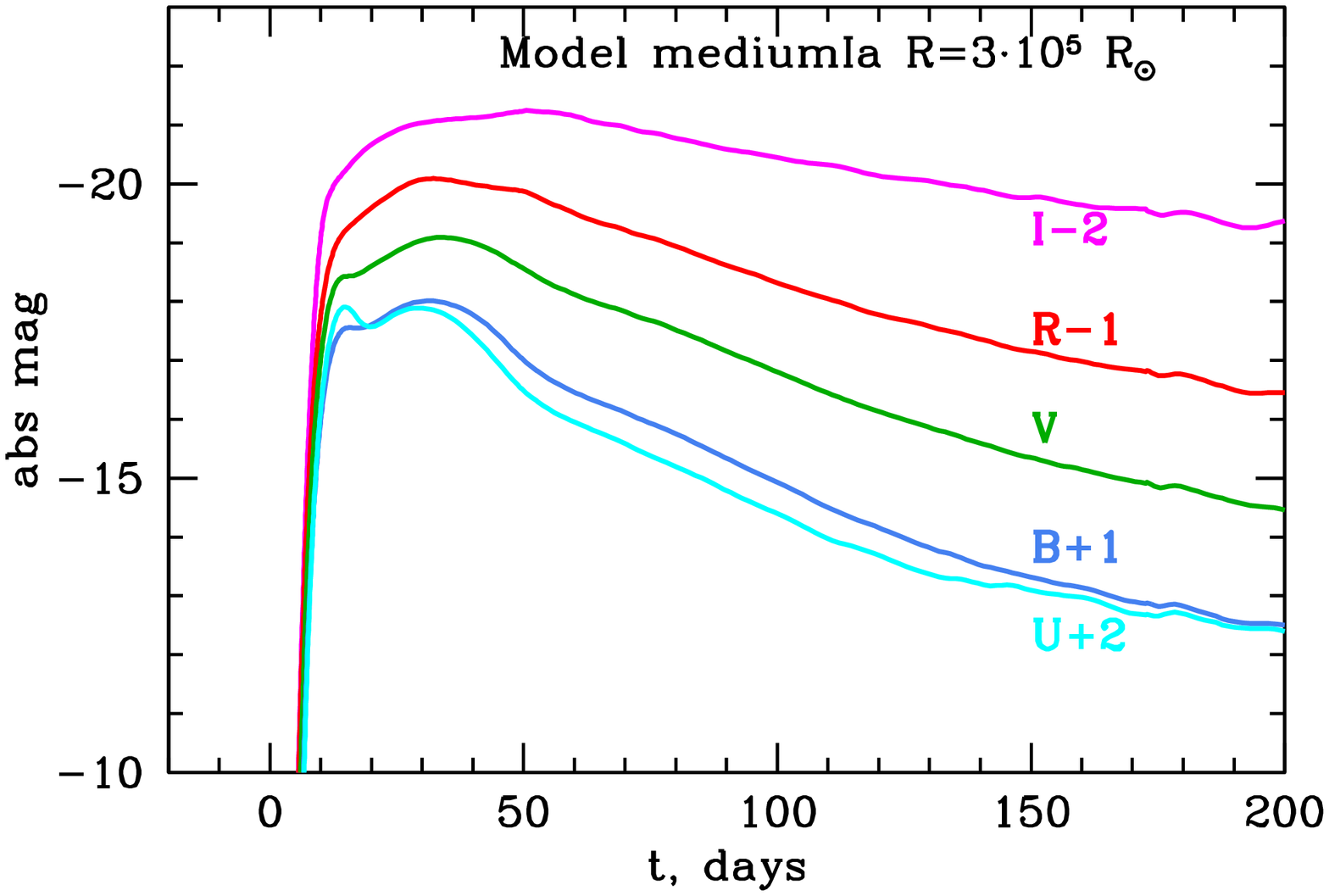}
\caption{\noindent Light curves
of the model {\tt mediumIa} with  $\rho(r)\propto r^{-3}$ in outer layers
and a smaller outer cut of the wind ($R=3\cdot 10^5 R_\odot$).
        }
\label{LCfryr3Rw5}
\end{figure}

The light curves for ``Ia'' runs are shown in Figs.~\ref{LCfryr4M2}--\ref{LCfryr3Rw5}.

The main effect which is clearly visible is the dependence
of the flux and duration of the light curve on the outer radius in the {\tt mediumIa}
(i.e. $\rho(r)\propto r^{-3}$) models.
A physical explanation for this may be a longer diffusion time in the larger
model.
However, we should investigate also the dependence of the results on grid resolution.
The presented ``Ia''  runs are done on grids with 200 radial by 100 frequency mesh points.
We have checked them on cruder grids (90 radial by 100 frequency meshes).
The behavior near maximum does not change much, while the tail part is more sensitive
to the grid.

A direct comparison of our results with those of \citet{Fryer2010} is not possible,
because we do not know their exact initial conditions.
Our {\tt mediumIa} models  ($\rho(r)\propto r^{-3}$  envelopes) have about twice as much
C--O as in \citet{Fryer2010} to exaggerate the effect of the circumstellar matter.
Still for a smaller outer radius of the wind our light curves seem to evolve
faster than in \citet{Fryer2010}.

One should also note that in Fig.~\ref{LCfryr3z2} we see initial ``plateau''
due to relatively high initial $T=2.5\cdot 10^3$~K in the ``wind''.
The radiation energy stored there is simply emitted away.
The process of formation of the extended structure may be much longer
than the time of its cooling.
It seems that at least part of the early light in Fig.~9 of \citet{Fryer2010} may be
attributed to this spurious emission.

\subsection{Type Ib explosions within C--O shells}
\label{Ib}

Supernova SN~2010gx \citep{Pastorello2010} is extremely luminous and
extremely interesting.
% One can seek its explanation on a way towards
For other very luminous events \citep{Ofek2007,Smith2007,Smith2010,GalMaz2009,Young2010}
some models involve explosions on a hypernova
scale, where  a huge amount of \nifsx is produced \citep[see, e.g., ][for type~IIn SN~2006gy
and type~Ic SN~2007bi]{NomTom2007,Moriya2010}.

Our goal is to explain SN~2010gx on another way, with minimum energy of explosion.
This seems to be possible based on the old idea due to \cite{GraNad1986}:
when we have two subsequent explosion events.
The first explosion is weak and produces the dense circumstellar structure
(which we call ``wind'' here, but which is not a steady wind, of course).
The second, normal SN explosion, produces very bright light due to the shock
embedded into a dense medium.
A physical mechanism for those multiple explosions,  proposed by \cite{HegWoo2002}
(pulsation pair instability) was used by \cite{WooBliHeg2007}  to explain
SN~2006gy with moderate energy of $\sim 3$~foe without any radioactive material.
One should keep in mind that there may be other, still unexplored, routes to
repeated explosions in stellar evolution, especially in binary systems.

It is interesting to look at the results of our ``Ib'' runs with \emph{ zero} \nifsx mass,
where one can see a pure effect of radiative shock in C--O envelope producing a very bright
supernova.

\begin{figure}
\centering
\includegraphics[width=0.45\linewidth]{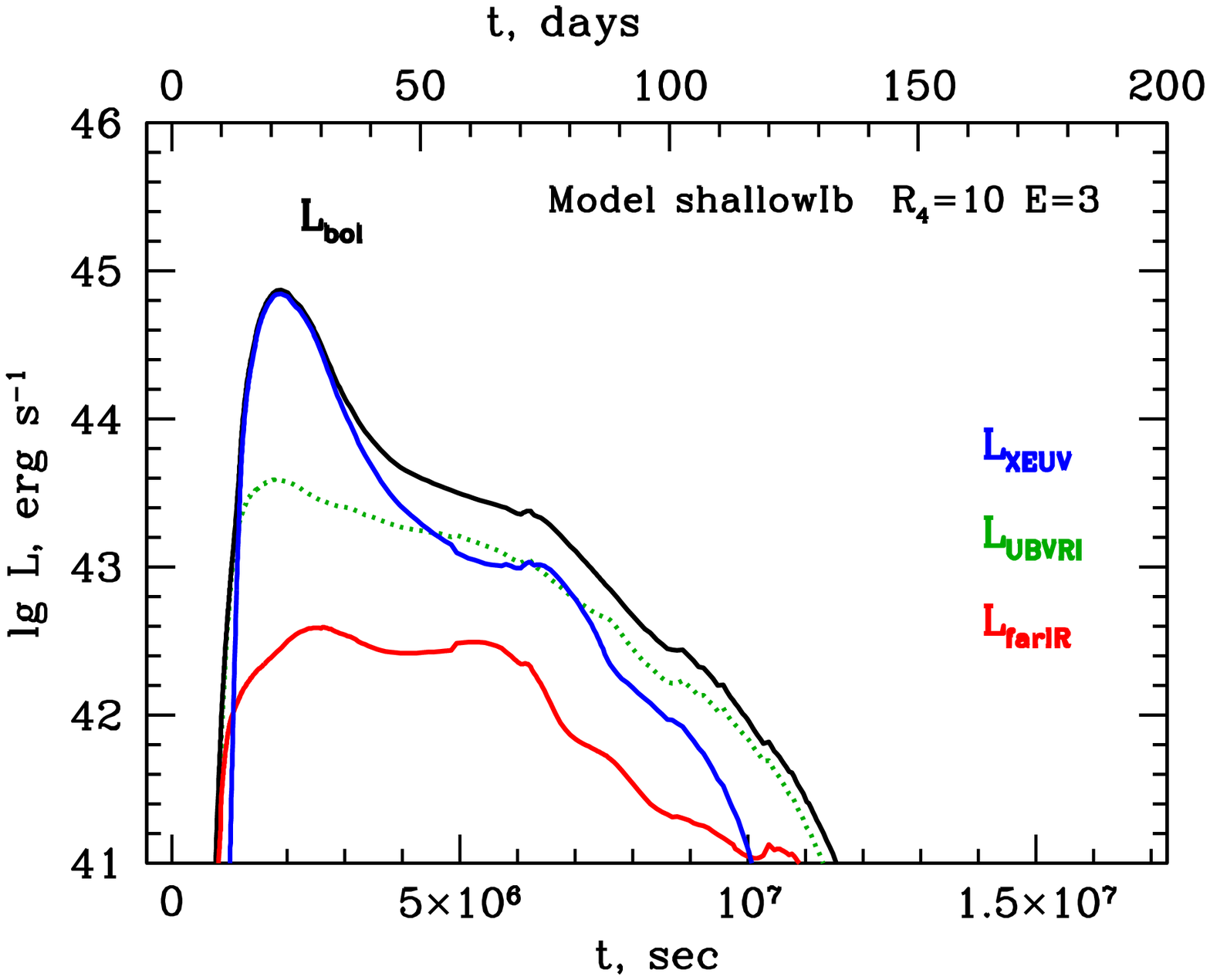}
\includegraphics[width=0.45\linewidth]{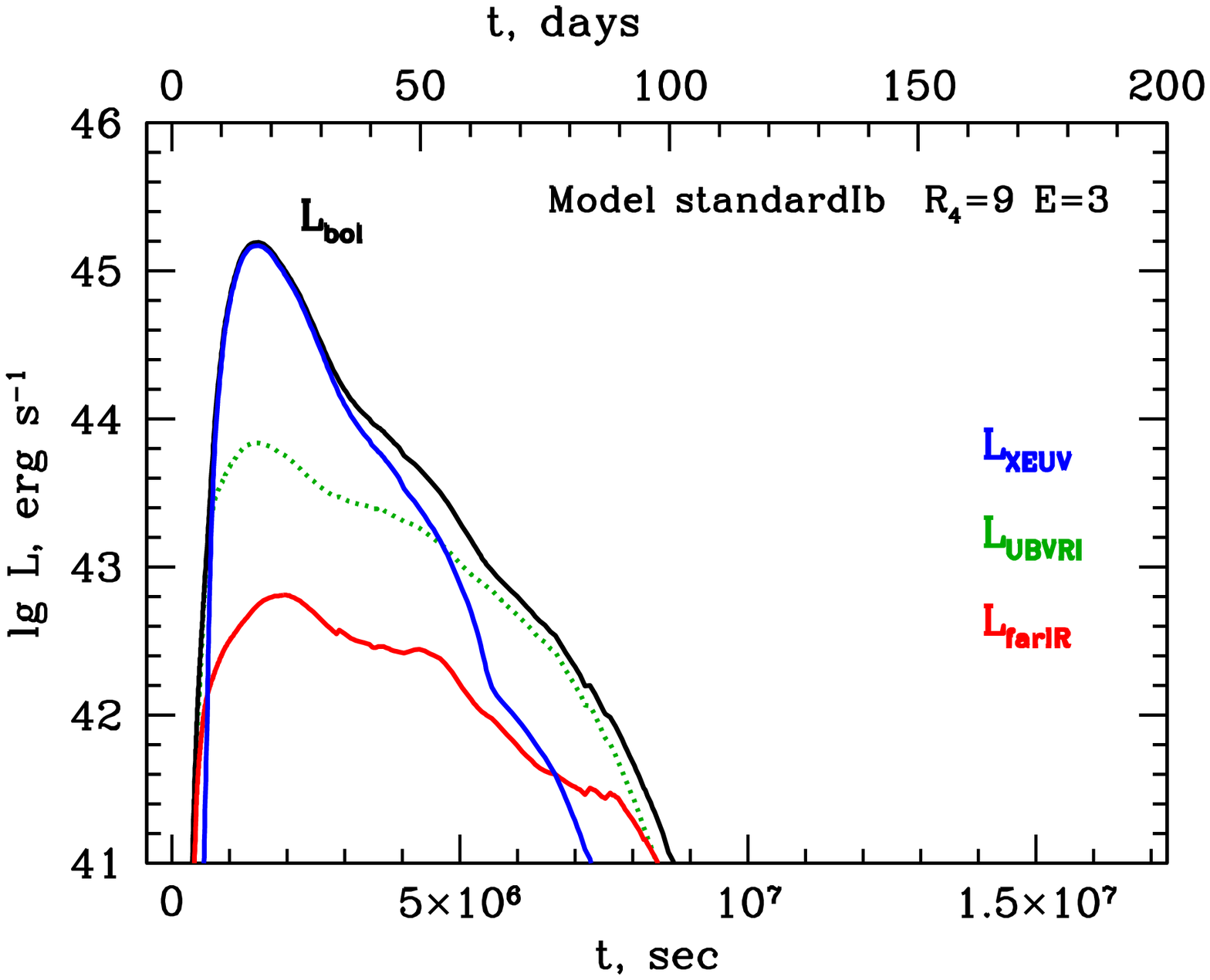}
\caption{\noindent Light curves
of the models {\tt shallowIb}  and  {\tt standardIb}.
        }
\label{LCshallowIb}
\end{figure}

One clearly needs a very large radius to produce a bright event.
The radius should not be too large, otherwise the diffusion of photons is
too slow and the light curve becomes too wide.
One needs high density for strong production of visible
light by the shock, but not too high, otherwise the mass of the ``wind''
and the optical depth of the shell become too large.

\begin{figure}
\centering
\includegraphics[width=0.45\linewidth]{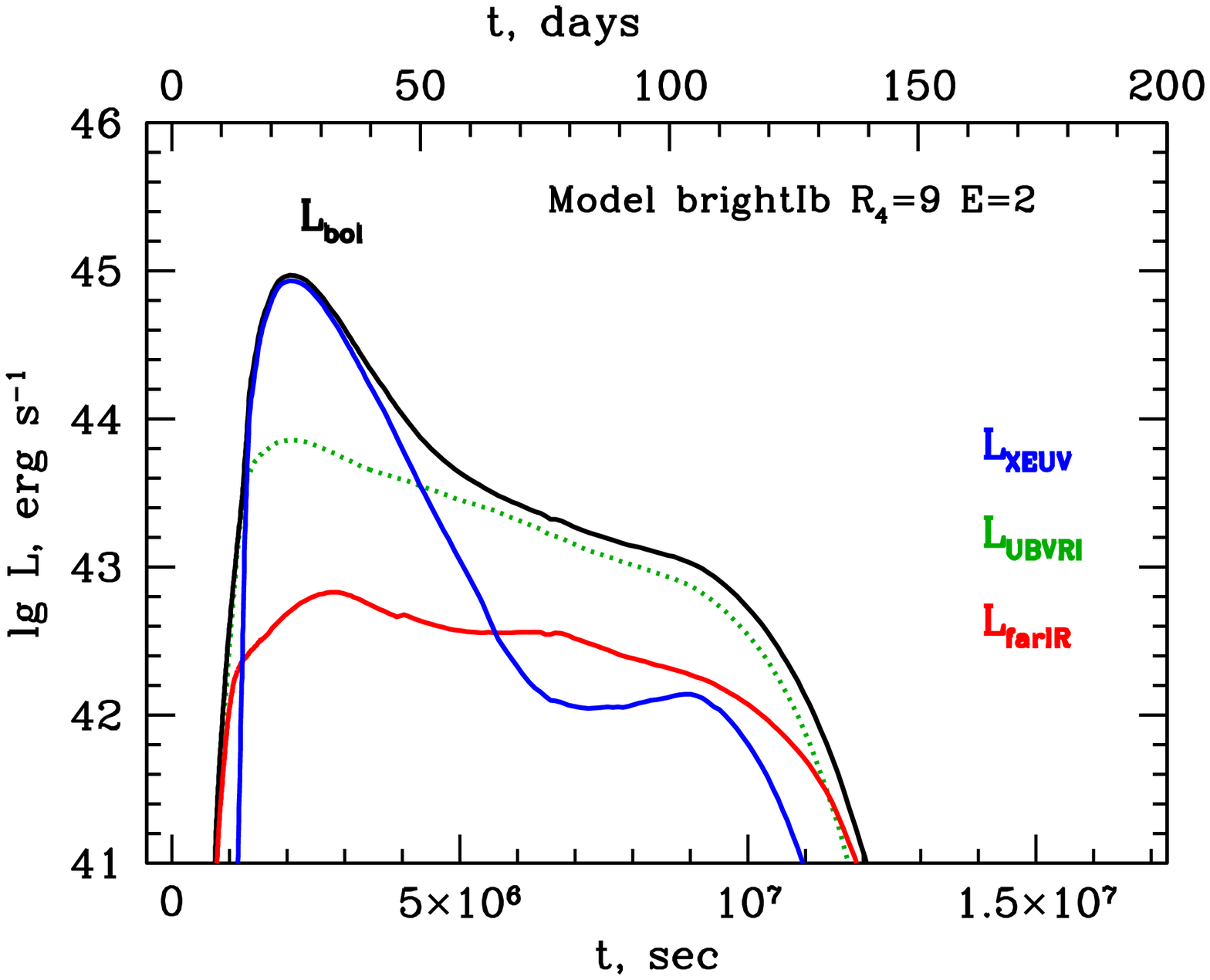}
\includegraphics[width=0.45\linewidth]{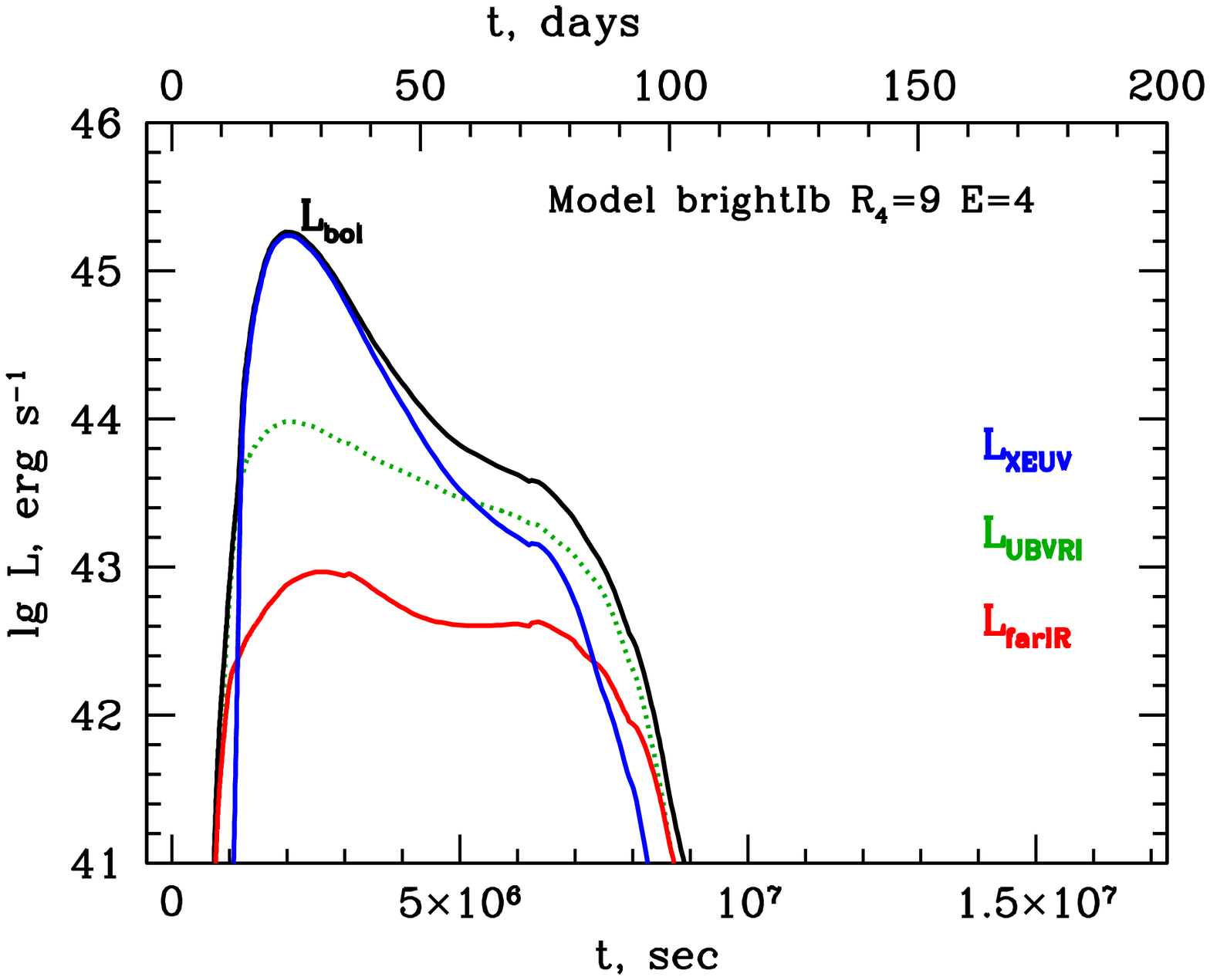}
\caption{\noindent Light curves
of models {\tt brightIb} with energies $E=2$~foe (left) and $E=4$~foe (right).
        }
\label{LCbrightIb}
\end{figure}

We have run tens of different models, varying all parameters.
Here we show only a small part of our results, see
Figs.~\ref{LCshallowIb}--\ref{LCbrightIb}.
Note, that the scale for the luminosity $L$ has changed in comparison to
Figs.~\ref{LCfryr4M2}--\ref{LCfryr3Rw5}.
The values of $p=4$ and $p=3$ in the $\rho(r)\propto r^{-p}$  distribution
produce too steep density gradients for our goal and we try smaller
values of $p$.
The models denoted {\tt shallowIb} have  $p=2.5$.
We call models {\tt standardIb} if they have  $p=2$ because this is
a standard value for a steady stellar wind, not for our set of models.
The best value we find is $p=1.8$ in the {\tt brightIb} set.

There is an indirect evidence that the density profiles around supernovae
having strong circumstellar interaction may be shallow indeed \citep{Prieto2007},
or do have rather complicated structure due to presupernova evolution \citep{Dwar2010}.

\subsection{Comparison with SN~2010gx. Role of opacity}
\label{sec:result-general}

Our code {\sc stella} allows us to produce the light curve
in different filter systems.
Here we present them in SDSS $ugri$  filters
using the same standards and transmission functions as we did in
\citet{Phillips2007}.
Following \cite{Pastorello2010} we assume redshift $z=0.23$ for SN~2010gx.
However, we take the distance modulus=40.28 which we find for this redshift
in standard cosmology with $H_0=71$~km/s/Mpc, $\Omega_m=0.27$, and $\Omega_\Lambda=0.73$.
\citet{Pastorello2010} do not give the distance modulus explicitly.
We can guess that they assume a larger distance, because they compare
the values of observed and absolute $g$ magnitudes for the host galaxy.
The larger distance
can be obtained in a flat universe only for very low
$H_0$ and $\Omega_m$, perhaps excluded by current observational data.

\begin{figure}
\centering
\includegraphics[width=0.45\linewidth]{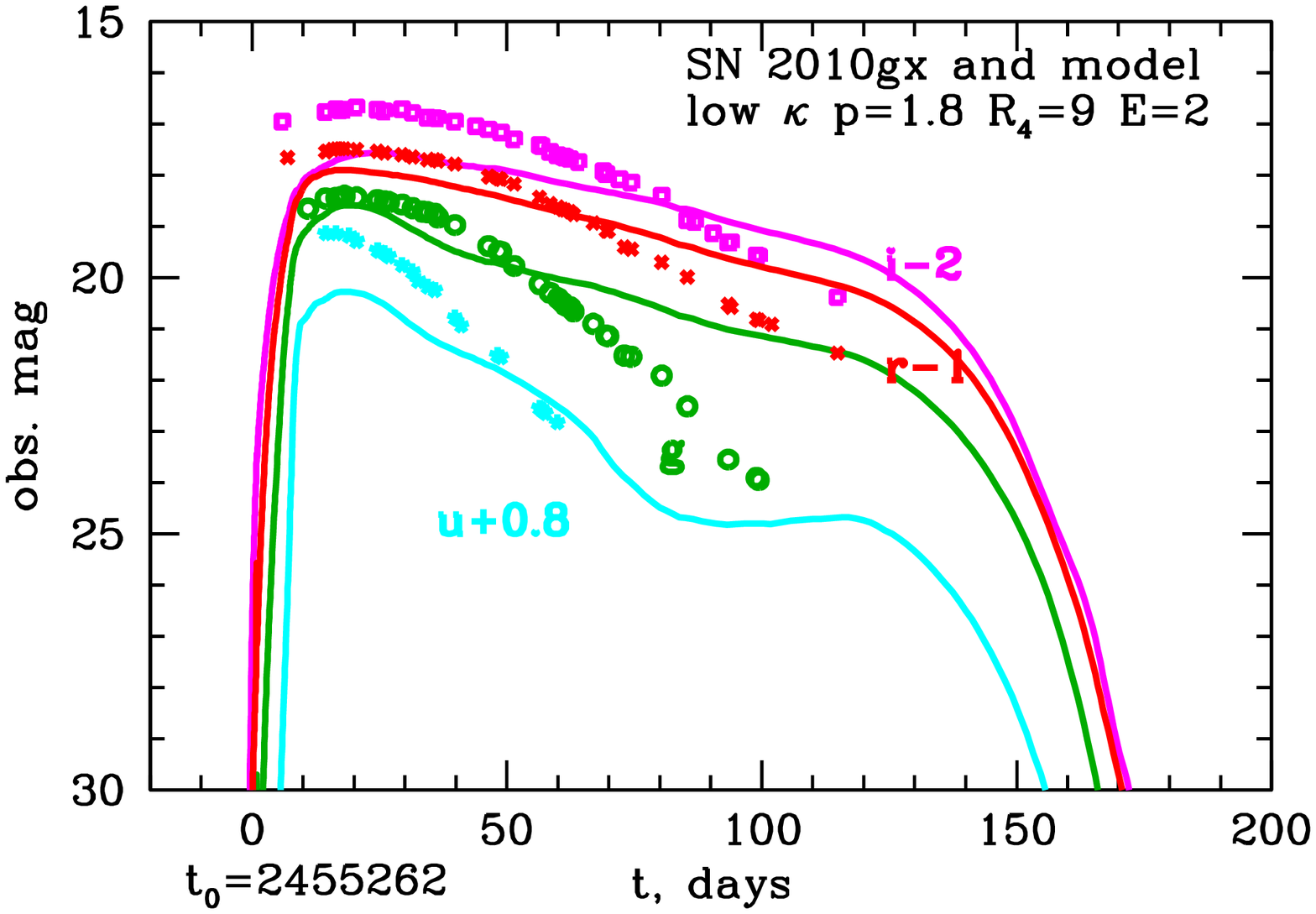}
\includegraphics[width=0.45\linewidth]{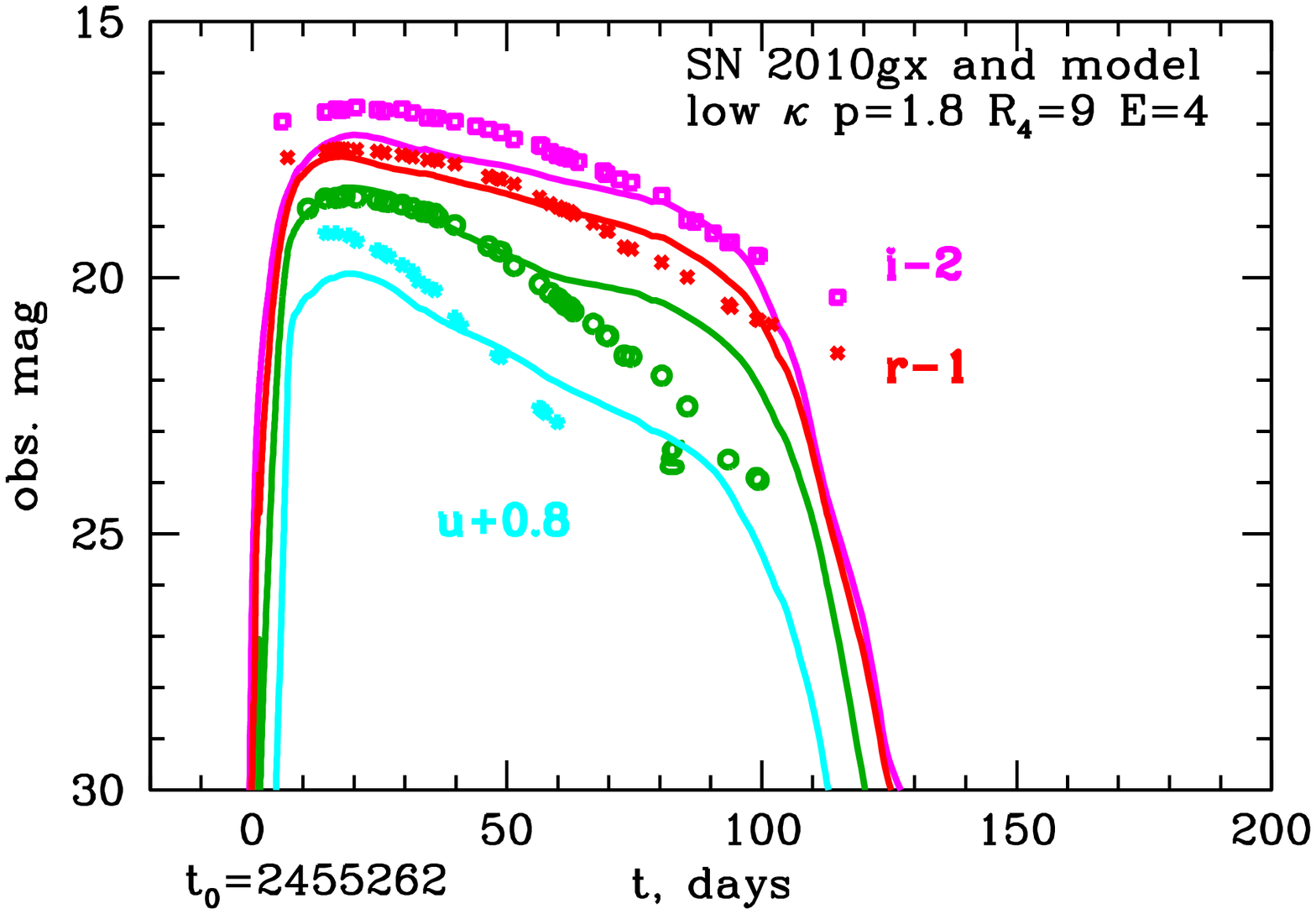}
\caption{\noindent Observed  and synthetic light curves in $ugri$ filters
of models {\tt brightIb} with energies $E=2$~foe (left) and $E=4$~foe (right).
        }
\label{zugribrightIb}
\end{figure}

Fig.~\ref{zugribrightIb} shows that our models {\tt brightIb} easily reach observed maximum
fluxes in $g$ filters for various values of explosion energy.
Energy $E=2$~foe is perhaps too low, energy $E=4$~foe is a bit too high, and
the decay of light is too fast.
The fluxes in $uri$ filters are a bit lower than in observations, but before
twiddling around those models we should note a very important factor, namely,
line opacity.

In our standard {\sc stella} setup we treat expansion opacity in lines as for type~Ia supernovae,
where we have homologous expansion and isotropic velocity gradient $dv/dr=1/t$, with $t$ time elapsed after the explosion.
Now we have $dv/dr \gg 1/t$, because the light is produced in the radiative shock attached
to a very dense shell (see Sec.~\ref{sec:hydroprofiles} below).
To check the effect of line opacity we have run several tests when
the expansion opacity is calculated with the fixed value $dv/dr=1\,\mbox{day}^{-1}$.
\begin{figure}
\centering
\includegraphics[width=0.45\linewidth]{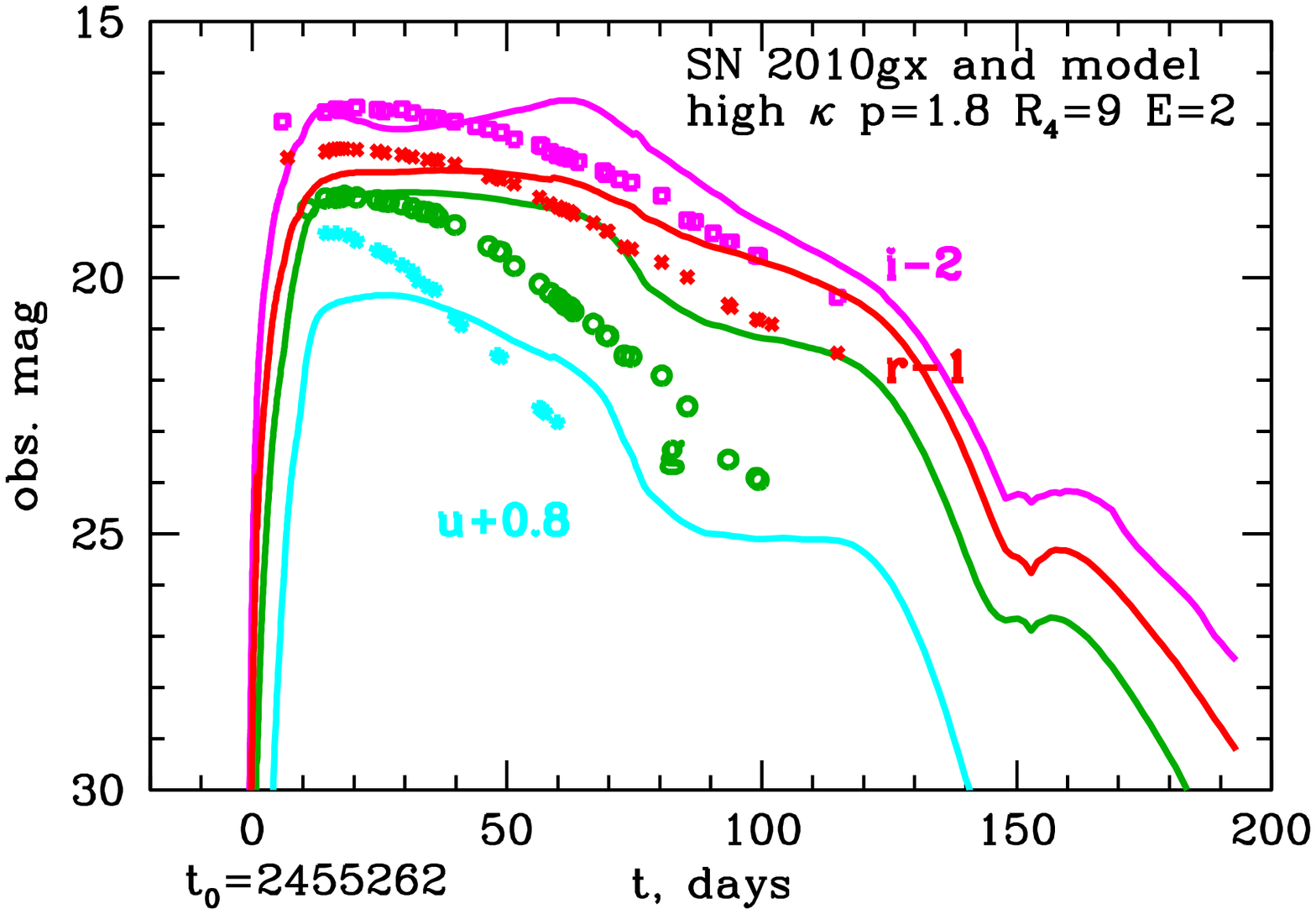}
\includegraphics[width=0.45\linewidth]{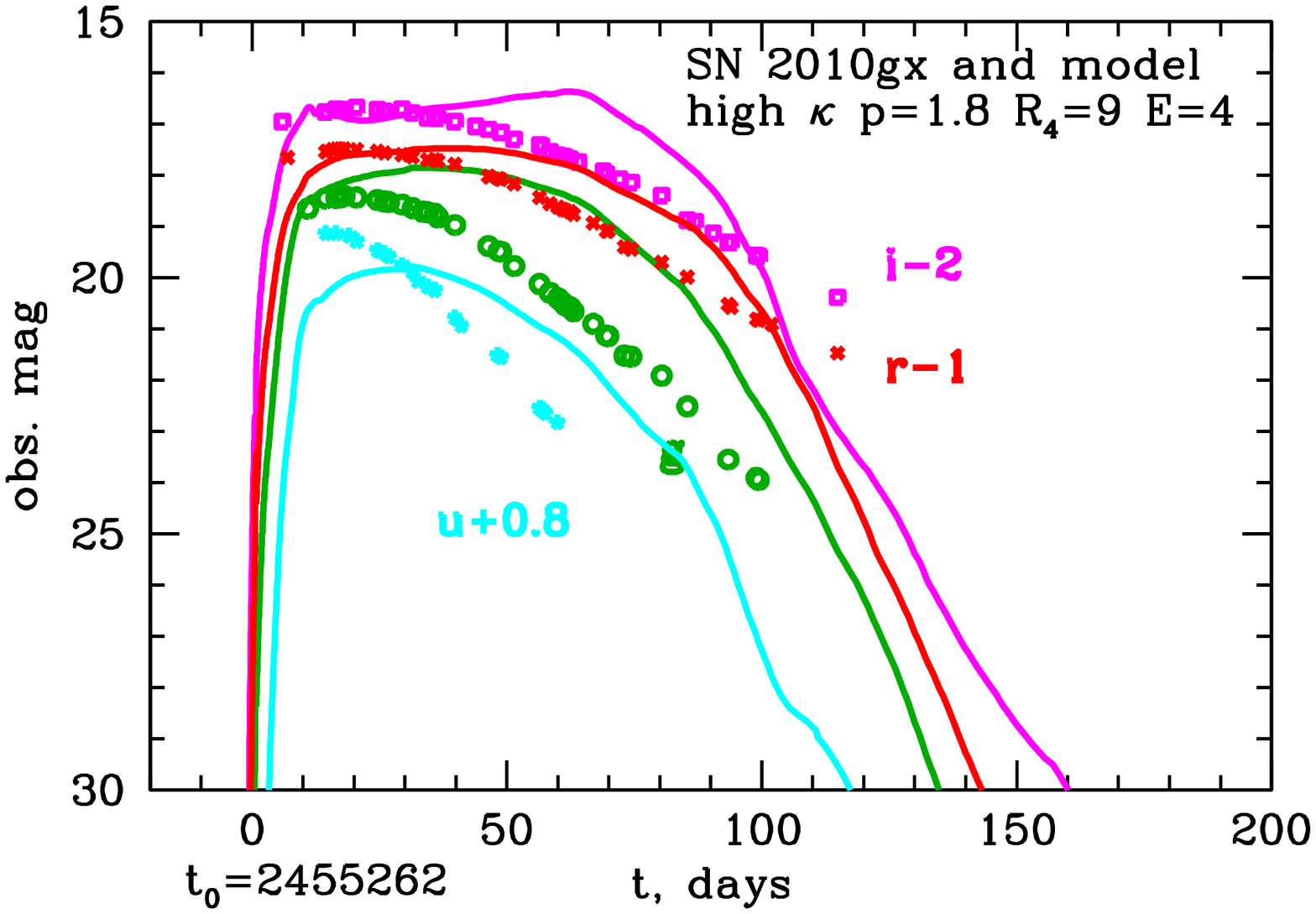}
\caption{\noindent Observed  and synthetic light curves in $ugri$ filters
of models {\tt brightIb} with energies $E=2$~foe (left) and $E=4$~foe (right)
for higher expansion effect in line opacity.
        }
\label{zugritf}
\end{figure}
Fig.~\ref{zugritf} shows that in this case the observed flux is higher in many bands.

\subsection{Hydrodynamic profiles of different models}
\label{sec:hydroprofiles}

\begin{figure}
\centering
\includegraphics[width=0.45\linewidth]{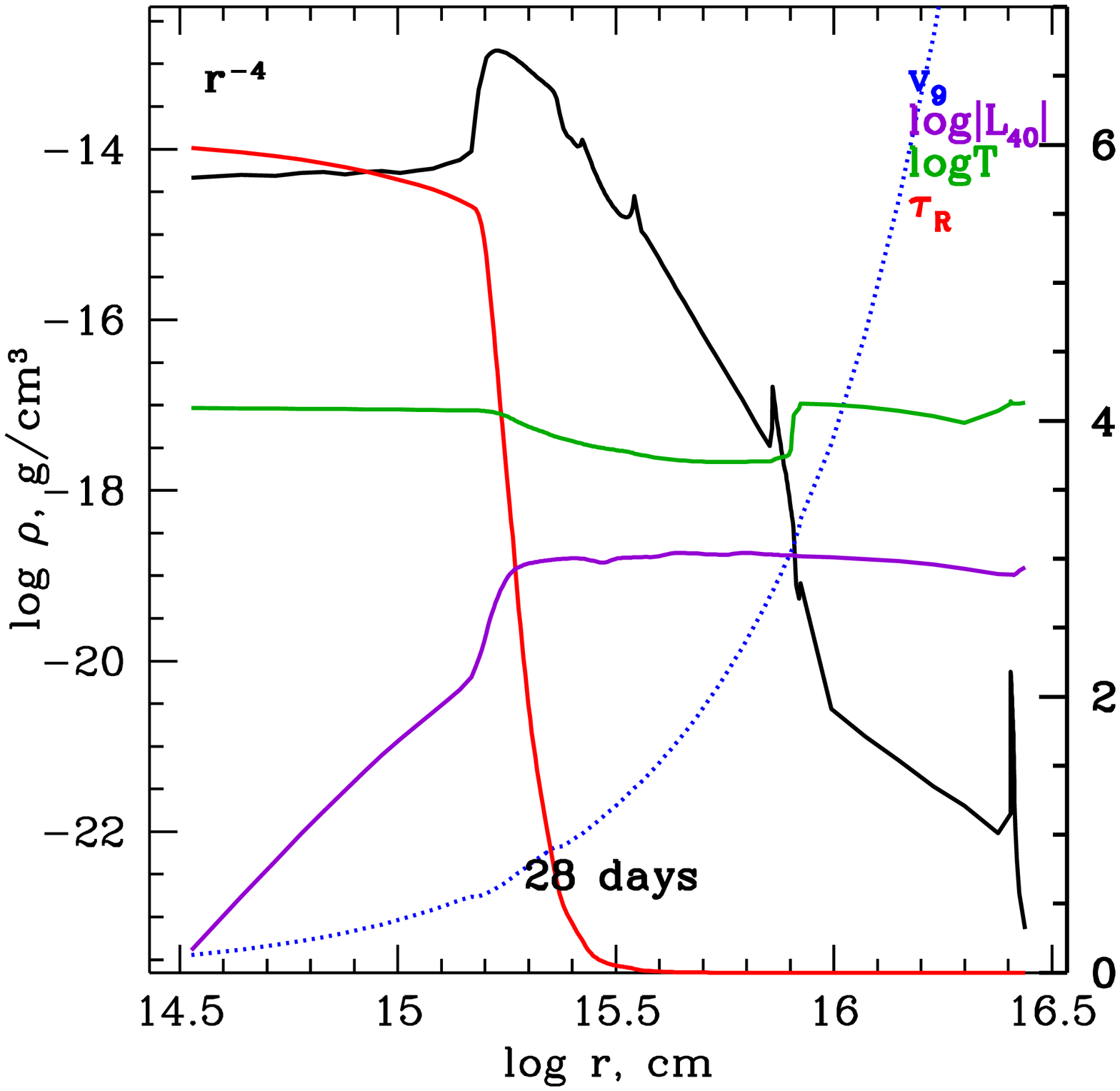}
\includegraphics[width=0.45\linewidth]{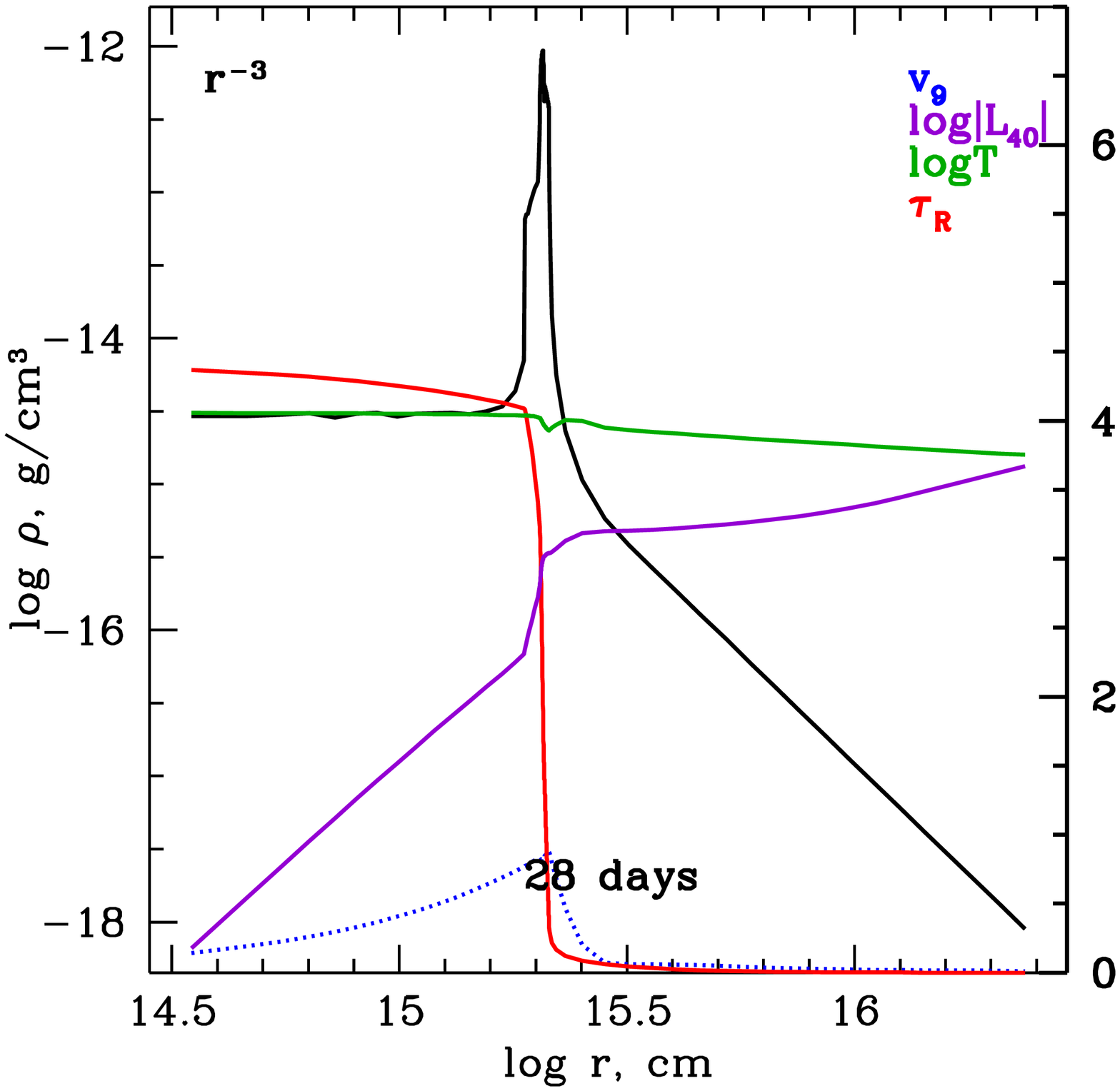}
\caption{\noindent Profiles of density (black lines),
velocity (in $10^9$~cm/s, blue), luminosity (logarithm of absolute value
in $10^{40}$~ergs/s, violet), logarithm of matter temperature (green) and Rosseland
optical depth (red). The scale for density is on the right Y axis,
for all other quantities it is on the left Y axis.
Left panel: model {\tt steepIa} with  $\rho(r)\propto r^{-4}$ in outer layers.
Right panel: model {\tt mediumIa} with  $\rho(r)\propto r^{-3}$ in outer layers
        }
\label{rhovr4M2}
\end{figure}

We present in Fig.~\ref{rhovr4M2} a comparison of details in distributions
of density, temperature, velocity, luminosity and Rosseland optical depth
in models {\tt steepIa} and {\tt mediumIa} for the same date a few weeks after
the explosion.

% \begin{figure}
% \centering
% \includegraphics[width=0.45\linewidth]{rhov28dLTtaulgr3Rw5}
% \includegraphics[width=0.45\linewidth]{rhov48dLTtaulgr3Rw5}
% \caption{\noindent Profiles of density (black lines), left Y axis,
% and velocity (in $10^9$~cm/s, blue), luminosity (logarithm of absolute value
% in $10^{40}$~ergs/s, violet), logarithm of matter temperature (green) and Rosseland
% optical depth (red) -- right Y axis,
% of the model with  $\rho(r)\propto r^{-3}$ in outer layers.
%         }
% \label{rhovr3Rw5}
% \end{figure}

We see that a shock wave in the  {\tt steepIa} model is much stronger
(due to a steeper density gradient) and leaves the grid quickly.
The luminosity is due partly to heat release from the shocked matter and partly to \nifsx,
but not to the shock itself: for the epoch shown in Fig.~\ref{rhovr4M2}
the shock has already left the grid.
Note that units of velocity on the plot are $10^9$~cm/s, i.e. ten thousand
km/s.
For a shallower profile $\rho(r)\propto r^{-3}$ the shock is ``buried'' deep
and forms a dense shell like in our SN~IIn models \citep{ChugaiEa04,WooBliHeg2007}.
One can see that the shock associated with this dense shell is
contributing to luminosity (although the  \nifsx role is not negligible).
It is interesting to study models with zero mass of \nifsx (as in \cite{WooBliHeg2007})
and we do this for SN~2010gx.

We do not observe a box-like profile of temperature as in Fig.10 of \citet{Fryer2010}.

\begin{figure}
\centering
\includegraphics[width=0.45\linewidth]{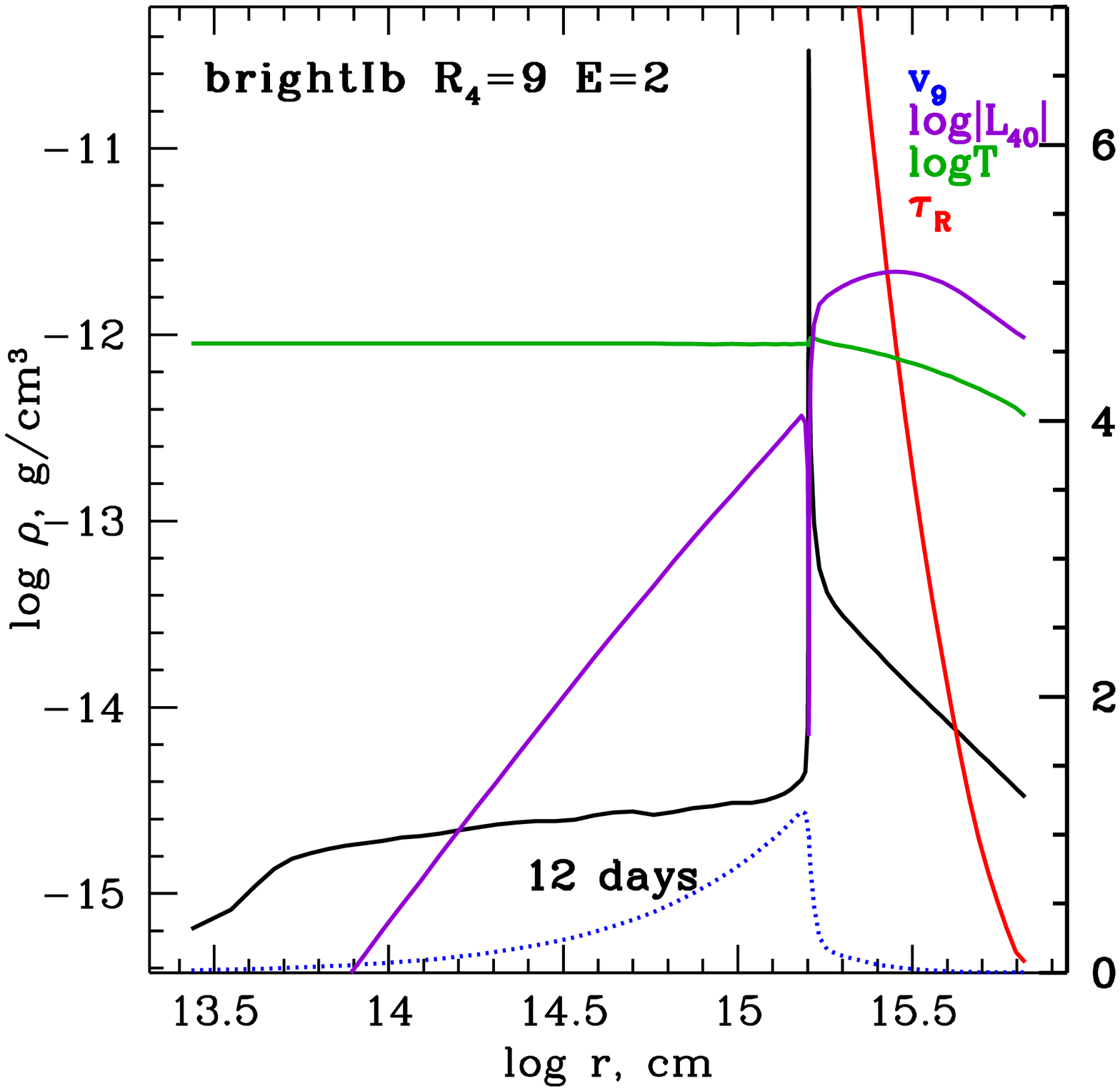}
\includegraphics[width=0.45\linewidth]{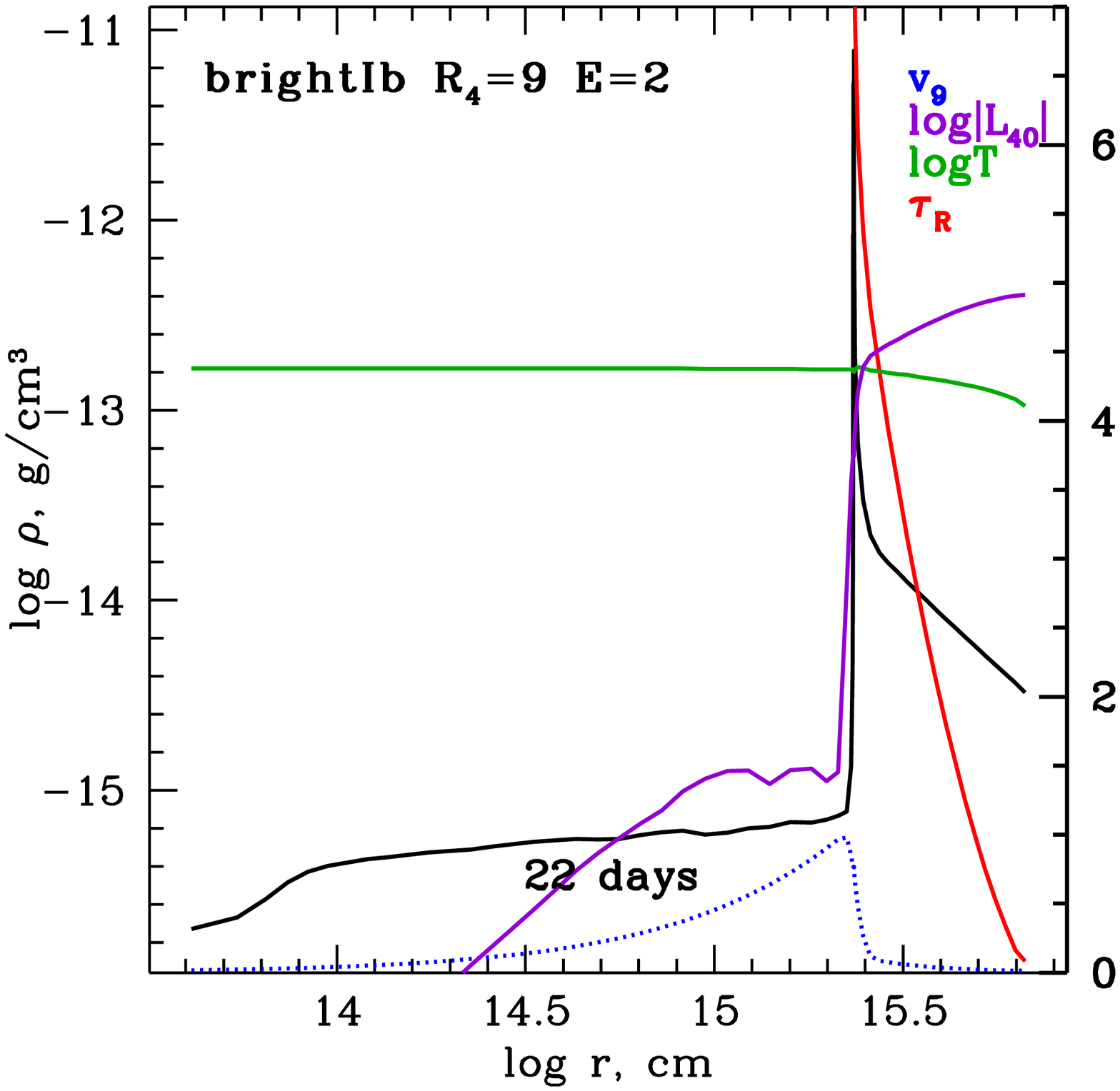}
\caption{\noindent The same as in Fig.~\protect\ref{rhovr4M2}, but
for the model {\tt brightIb} with  $\rho(r)\propto r^{-1.8}$ in outer layers and $E=2$~foe.
Left panel: at day 12.
Right panel: at day 22.
        }
\label{rhovrbright}
\end{figure}

Fig.\ref{rhovrbright} shows profiles for one of our models of SN~2010gx.
We see a formation of very dens shell and luminosity production on the shock.

\begin{figure}
\centering
\includegraphics[width=0.45\linewidth]{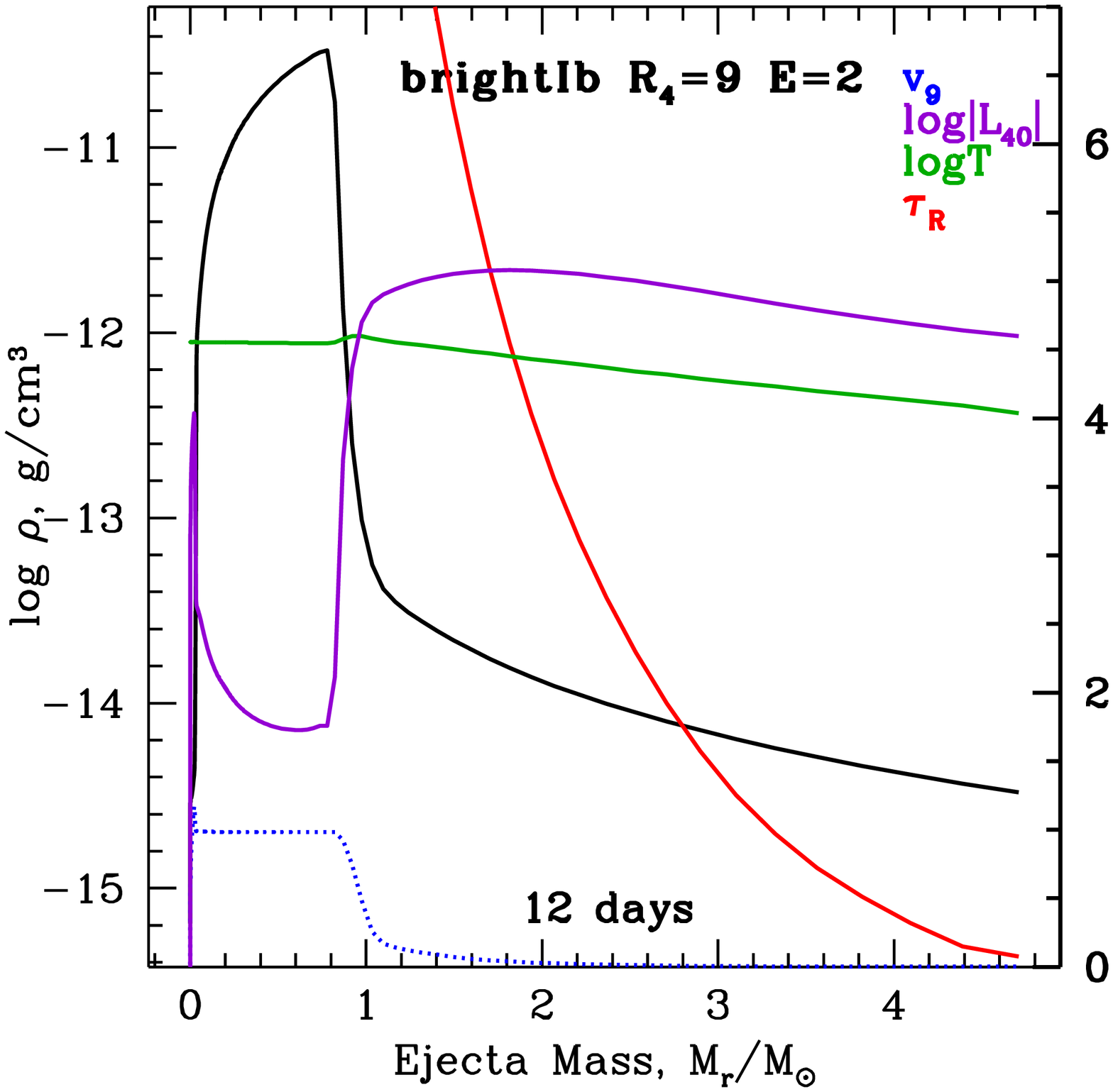}
\includegraphics[width=0.45\linewidth]{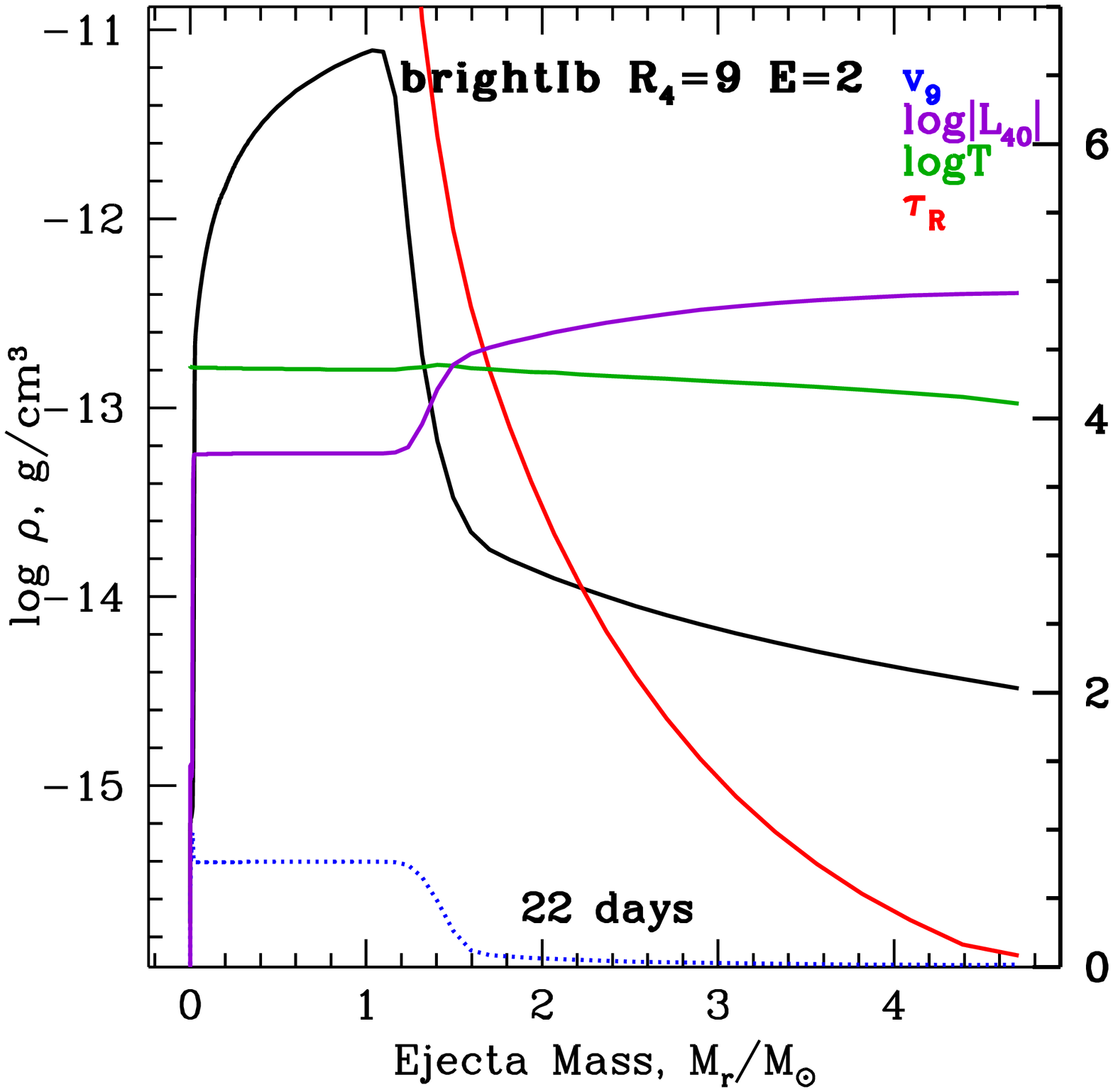}
\caption{\noindent The same as in Fig.~\protect\ref{rhovrbright}, but
as a function of $M_r$.
% Profiles of density (black lines), left Y axis,
% and velocity (in $10^9$~cm/s, blue), luminosity (logarithm of absolute value
% in $10^{40}$~ergs/s, violet), logarithm of matter temperature (green) and Rosseland
% optical depth (red) -- right Y axis.
Left:
% model {\tt brightIb} with  $\rho(r)\propto r^{-1.8}$ in outer layers and $E=2$~foe
at day 12.
Right: at day 22.
        }
\label{rhovmbright}
\end{figure}

\section{Summary and discussions}
\label{sec:summary}

The aim of the current work was to check numerically the role of shock
interaction in producing light by shocks in carbon-oxygen ``winds''
around type~I supernovae.
The problem arises from observations \citep[and references therein]{Pastorello2010},
as well as from theory \citep{Fryer2010}.
Our main conclusion is that the shock interaction is very important indeed and it cannot
be ignored in many cases.

We do not know exact initial conditions used by \cite{Fryer2010},
and we do not confirm their conclusion on long-living shocks in steep
($\rho(r)\propto r^{-4}$) density profiles.
We suspect that the shock lives there due to another distribution of matter
(there is perhaps a medium with $\rho(r)=$~const on their grid).
In part of our results we see initial flashes of light due to initial temperature
in the ``wind'', even if it is as low as 2500~K.
Nevertheless, we find long-living radiating shocks in all profiles with
$\rho(r)\propto r^{-3}$ and in the less steep ones.

A big question remains %to \citet{Fryer2010}
on the mechanism of the formation of those huge and dense envelopes.
What is the time-scale of this process?
How far can the  envelope extend in radius? What
is the density profile and the temperature of matter before the core explodes?

If those structures form in reality then we
can have extremely bright and long light curves.
This question deserves further investigation.
The effect of the extended envelopes is probably more important not for SNe~Ia
but for very luminous SNe~Ib/c discovered recently \citep{Pastorello2010}.
In that case one may think about pulsational pair instability as a means
for formation of extended shell.
One can also speculate about mergers of white dwarfs with CO-cores of WR stars.
The merger event may lead to a moderate explosion with energy of a few percent of
foe which forms a cloud of matter around the presupernova expanding
with relatively low velocity.
Then, if the core collapse follows within \emph{ years} after the merger,
one can have all needed conditions for very bright supernovae.

Our simulations show that the very luminous SN~2010gx
may indeed be produced by a supernova explosion if there is enough surrounding
material for a shock to transform the kinetic energy of ejecta
into observed light.
We find that SN~2010gx can be
explained at explosion energies $\sim (2 \div 3)$~foe for a non-steep density
profile, if the total mass of SN ejecta and a shell is  $\sim (3 \div 5)\, M_\odot $
and the radius of the shell is  $\sim 10^{16} $~cm.

The fits to fluxes in individual filters are not yet perfect in our simulations.
This is natural: we have taken quite arbitrary and primitive
chemical compositions, density distributions, etc.
But in many cases the synthetic fluxes are \emph{higher} than observations,
and this looks encouraging.
One can try building a better fit to observations by variations of initial
conditions in the model.
However, it seems that it is to early to optimize the models along these lines.
This optimization will probably not give us a true insight and
a better understanding of the problem.

E.g., many technical subtleties still remain in the treatment of line opacity.
First, we should correct for the expansion effect not only the flux equation,
but also the energy equation.
Microscopically, for exact monochromatic opacity, there is
no expansion effect \citep{Bli1996,Bli1997}.
For average absorption opacity there is some effect, but it
can not be treated so simply as by  \cite{Hoeflich1993},
\citep[see][]{SorBli2002}.
Moreover, there is another complication with anisotropic velocity
gradient.
Thus, before optimizing the fits, one has to build new techniques
for radiation transfer in these conditions.

We have not discussed observed line spectra:
for type IIn, by definition, one clearly sees narrow emission lines produced in the shells.
Not so in SN~2010gx: narrow circumstellar lines are not seen \citep{Pastorello2010}.
There is no hydrogen, which is easily excited, and the most abundant elements,
carbon and oxygen, should be present perhaps as C~II and O~II ions in the envelope
(see the temperature in Figs.~\ref{rhovrbright},\ref{rhovmbright}).
These ions do not have many strong lines in visible light.
It is not easy to identify C and O lines on photospheric stage in SNe~Ic \citep{Young2010},
and now they should be excited even if there is no
radioactive material.
So one has to look for \emph{weak and narrow} lines in noisy spectra.
This problem certainly deserves further investigation with account
of different conditions for ionization/excitation of shells under the shock radiation.

The main complication to the whole picture is possible fragmentation of the dense shell.
The attempts on multi-D treatment of SN ejecta evolution are rather old
\citep{TenT1991,CheBlo1995,BloLunChe1996}, more recent
results and references may be found in \citep{Dwar2007,Dwar2008}.
See also \citet{VMarle2010} for the case of SN~2006gy, but without real treatment of radiative
transfer.
There are several 3-D MC transport codes
\citep{Hoeflich2002,Lucy2005,Kasen2006,Kasen2007,Sim2007,Tanaka2008,Kromer2009}
but they are not actually coupled to hydrodynamics and there are many
difficulties in doing this \citep{Almgren2010}.

Full NLTE treatment is needed to predict spectra,
but very little is done on this even for SN~IIn.
E.g., \citet{Dessart2009} are surprisingly successful in reproducing the spectra
of SN~1994W in a set of atmospheric models, but their method is applicable
only to monotonic velocity structures, not to shocked shells.
Moreover, one should be cautioned about the relation of ``photospheric''
radius found by \citet{Dessart2009} which shrinks,
and the radius of the shocked shell in SNe~IIn which grows.
This is already explained by \citet{SmithChor2008,Smith2010}.

Nevertheless, we conclude that, provided the formation of rather dense and extended
circumstellar shells, the extremely powerful events of type~Ib/c
like SN~2010gx can be explained with moderate energy of explosions
without invoking any radioactive material.

\section*{Acknowledgements}

This work was begun while we %(S.~Blinnikov and E.~Sorokina)
were visiting Max Planck Institut f{\"u}r Astrophysik, Garching, in July--August, 2010.
We are especially grateful to Wolfgang~Hillebrandt,  to Stuart~Sim
for drawing our attention to the results of Fryer et al., and to Ashly Ruiter
for discussions on possible channels in binary stellar evolution.
SB is thankful to Victor Utrobin and to Takashi~Moriya for discussions on bright SNe~Ic
spectra, light curves, and
on numerical modelling of radiation dominated shocks with our code
{\sc stella}.

Our work in Germany is supported by MPA guest program, and in Russia partly
by grants RFBR 10-02-00249-a,
10-02-01398-a, by Agency for Science and Innovations contract 02.740.11.0250,
by Sci.~Schools-3458.2010.2, % Bisnov
-3899.2010.2 % Sharkov
 and by the grant IZ73Z0-128180/1 of the Swiss National Science
Foundation (SCOPES).

% РФФИ: 10-02-00249-а, 10-02-01398-а,
% НШ 2977.2008.2, 3884.2008.2,
% Агентство по науке и инновациям - контракт: 02.740.11.0250 (НОЦ)
% грант SNSF (Swiss National Science Foundation)
% по программе  SCOPES  No.~IZ73Z0-128180/1

\bigskip

%{\bf References}
\bibliographystyle{aa} % style aa.bst
\bibliography{cowindsne} % your references Yourfile.bib

%\begin{thebibliography}
%\end{thebibliography}

\newpage

\end{document}